\documentclass[final]{article}
\usepackage[final]{graphicx}
\usepackage{graphicx}
\usepackage{amssymb}
\usepackage{amsfonts}
\usepackage{amsmath}
\usepackage{rotating}
\usepackage{pdflscape}
\usepackage[english]{babel}
\usepackage{color}
\usepackage{bigintcalc}
\usepackage{bbm}
\usepackage{tabularx}
\newcolumntype{L}{>{\raggedright\arraybackslash}X}
\usepackage{listings}
\usepackage{numprint}
\usepackage{bigintcalc}
\usepackage{seqsplit}

\usepackage[utf8]{inputenc}
\usepackage[T1]{fontenc}
\usepackage{listings}
\usepackage{xcolor}
 
\definecolor{codegreen}{rgb}{0,0.6,0}
\definecolor{codegray}{rgb}{0.5,0.5,0.5}
\definecolor{codepurple}{rgb}{0.58,0,0.82}
\definecolor{codeblue}{rgb}{0.1,0.1,0.8}
\definecolor{backcolour}{rgb}{0.95,0.95,0.92}
 
\lstdefinestyle{mystyle}{
    backgroundcolor=\color{backcolour},   
    commentstyle=\color{codegreen},
    keywordstyle=\color{codeblue},
    numberstyle=\tiny\color{codegray},
    stringstyle=\color{codeblue},
    basicstyle=\footnotesize,
    breakatwhitespace=false,         
    breaklines=true,                 
    captionpos=b,                    
    keepspaces=true,                 
    numbers=left,                    
    numbersep=5pt,                  
    showspaces=false,                
    showstringspaces=false,
    showtabs=false,                  
    tabsize=2
}
 
\begin{document}
\lstset{language=Mathematica}

\title{Computing the solutions of the van der Pol equation to arbitrary precision}

\author{
Paolo Amore \\
Facultad de Ciencias, CUICBAS, Universidad de Colima,\\
Bernal D\'{i}az del Castillo 340, Colima, Colima, Mexico  \\
paolo@ucol.mx}
\maketitle

\begin{abstract}
We describe an extension of the Taylor method for the numerical solution of ODEs that uses Pad\'e approximants to obtain extremely precise numerical results.
The accuracy of the results is essentially limited only by the computer time and memory, provided that one works in arbitrary precision. In this method the stepsize 
is adjusted to achieve the desired accuracy (variable stepsize), while the order of the Taylor expansion can be either fixed or changed at each iteration (variable order).

As an application, we have calculated the periodic solutions (limit cycle) of the van der Pol equation with an unprecedented accuracy for a large set of couplings (well beyond the values currently found in the literature) and we have used these numerical results to validate the asymptotic behavior of the period, of the amplitude and of the Lyapunov exponent reported in the literature.
We have also used the numerical results to infer the formulas for the asymptotic behavior of the fast component of the period and of the maximum velocity, which have never been calculated before.
\end{abstract}

\section{Introduction}
\label{Intro}

The Taylor method is one of the oldest methods for the numerical solution of ODEs: in fact 
the well--known Euler method is essentially  a Taylor method of order one. 
Higher order Taylor methods have been explored in the literature, but their use is still not widespread. 
From our (partial) exploration of the literature, we have found refs.~\cite{Deprit66, Corliss82, Chang94, Jorba05, Barrio05, Barrio05b, Barrio11, Barrio12}
to be the most relevant to our discussion.

The paper by Deprit and Zahar~\cite{Deprit66} is possibly the first to show the advantages of using the Taylor method compared to a 
more standard Runge-Kutta-Nystrom algorithm: these authors found that the latter needs to use step size about $4000$ smaller to reach the same accuracy of the former.
Corliss and Chang~\cite{Corliss82,Chang94} have applied the Taylor method to a series of problems (using fortran). 
The remaining papers, ref.~\cite{Jorba05} by Jorba and Zou and refs.~\cite{Barrio05, Barrio05b, Barrio11, Barrio12} by Barrio and collaborators, 
are particularly important because of their emphasis on achieving high precision. In general it is found that  the high-order Taylor method is very competitive
with other numerical methods when high accuracy is needed; moreover this method can be used to calculate the numerical solutions to certain class of ODEs with unprecedented accuracy (from the abstract of ref.~\cite{Barrio11}: "this methodology is the only one capable of reaching precision up to thousands of digits for ODEs").

At the same time, Jorba and Zou clearly state a limitation to the Taylor method: "The main drawback is that
the Taylor method is an explicit method, so it has all the limitations of these kind of schemes. For instance, it is not suitable for
stiff systems.". A discussion of the applicability of Taylor method for stiff ODEs is also contained in \cite{Barrio05}~\footnote{One of the examples considered in that paper is the van der Pol equation; in this case the authors state:"If the parameter is large, the differential system becomes stiff, and we will need implicit methods
in its solution." }.

The main goal of our paper is to describe an extension of the Taylor method that allows to obtain arbitrarily accurate numerical solutions even for stiff differential equations, such as the van der Pol equation that will be used in this paper as the ideal testing ground. The method that we propose is based on the use of Pad\'e approximants rather than Taylor polynomials; we will show that the performance of the standard Taylor method is drastically improved both by enlarging the class of problems that can be attacked (e.g. stiff ODEs) and also by increasing the accuracy of the solutions with respect to  the Taylor method with the same number of coefficients.

We will illustrate our method by applying it to the van der Pol oscillator, the most famous example of ODE supporting a limit cycle. 
For this problem there is an extensive literature, but surprisingly enough numerical solutions have never been calculated to 
a high level of accuracy, particularly for the stiff regime. Moreover the behavior of the period and amplitude of the limit cycle has been studied in depth both for small coupling (using perturbation theory, i.e. the Lindstedt-Poincar\'e method) and for large coupling (using asymptotic methods): by producing extremely accurate numerical solutions for a wide range of couplings we plan to show that it is possible to reproduce the known analytical results, but also to  predict the asymptotic behavior of quantities not yet studied.

In our opinion, the quest for extremely precise numerical solutions is not a mere exercise of virtuosity but it has real  applications: one cannot overstate the importance of predicting the behavior of a dynamical system while keeping under control the precision of the results. We believe that the technique described in this paper may be applied to a large number of problems for which accuracy is important.

The paper is organized as follows: in Section \ref{sec:method} we introduce our Pad\'e -- Taylor method (PTM) and discuss its main features; in Section \ref{sec:vdP} we consider the van der Pol oscillator and discuss the application of the Pad\'e--Taylor method to it; 
in the subsections \ref{sec:num} and \ref{sec:extra} we present the numerical results and apply appropriate extrapolation method to extract the leading asymptotic behavior for different quantities (particularly the period and the amplitude of the oscillator); finally, in Section \ref{sec:concl} we draw our conclusions and  discuss the possible directions of future work.

\section{The Pad\'e -- Taylor method}
\label{sec:method}

Consider the ODE
\begin{equation}
\frac{d{\bf X}}{dt} = F(t, {\bf X}(t)) \hspace{1cm} , \hspace{.2cm} {\bf X} \in \mathbb{R}^d   \hspace{.2cm} , \hspace{.2cm} t \in \mathbb{R}
\label{eq:ode}
\end{equation}
with initial conditions
\begin{equation}
{\bf X}(t_0) = {\bf X}_0 
\end{equation}

At $t>t_0$ the solution of eq.~(\ref{eq:ode}) can be approximated by a Taylor polynomial of order $N$
\begin{equation}
{\bf X}(t) \approx {\bf X}(t_0) + \left. \frac{d{\bf X}}{dt} \right|_{t_0} (t-t_0) + \dots + 
\frac{1}{N!} \left. \frac{d^N{\bf X}}{dt^N} \right|_{t_0} (t-t_0)^N
\label{eq:Taylor}
\end{equation}

The Taylor coefficients in eq.~(\ref{eq:Taylor}) can be obtained by repeated derivations of eq.~(\ref{eq:ode}) and by subsequently setting $t=t_0$. Let $T_0$ be the radius of convergence of the power series obtained by taking the limit $N\rightarrow \infty$: clearly we are justified to use the polynomial (\ref{eq:Taylor}) only for $t_0 < t \ll t_0+T_0$. Without going to the extreme case where the power series is divergent ($T_0=0$), the reader may recognize that a finite value of $T_0$ sets an upper limit to the stepsize in the numerical integration of eq.~(\ref{eq:ode}).
As a time step $\tau_0 \ll T_0$ is chosen (in some way -- we will discuss this point soon), a new approximation to the function at $t_1 \equiv t_0 +\tau_0$: from this approximate solution one can estimate the new Taylor coefficients corresponding to eq.~(\ref{eq:Taylor}) expanded arout $t_1$ and repeat the same arguments. In general we will have $\tau_1 \neq \tau_0$ (variable stepsize): additionally, at each step one can use polynomials of different order if needed (variable order).

Of course we run into trouble if the power series becomes divergent (vanishing radius of convergence) at any finite time $t_i$: in this case $\tau_i = 0$ is the only allowed stepsize.

The natural choice to extend the range of applicability of the power expansion is to substitute the Taylor expansion
with a Pad\'e approximant of appropriate order:
\begin{equation}
\mathcal{P}\left[{\bf X}(t)\right]_{[m,n]} = \frac{P_m(t-t_0)}{Q_n(t-t_0)}
\label{eq:Pade}
\end{equation}
where $P_m(t-t_0)$ and $Q_n(t-t_0)$ are polynomials of order $m$ and $n$ and the expansion of eq.~(\ref{eq:Pade}) reproduces the first 
$N+1$ Taylor coefficients ($m+n=N$).
Let $\tau_i^{(0)}$, with $i=1,\dots,n$, be the (complex) roots of the polynomial $Q_n(\tau)$: the smallest positive real root (if any) will limit the region of applicability  of (\ref{eq:Pade}) to smaller values of the stepsize.

Let us now how come to the important question: how do we select a suitable stepsize $\tau_0$? The only limitations on $\tau_0$ that we have discussed so far are related to the radius of convergence of the power series at $t=t_0$ and to the possible presence of 
spurious singularities of the Pad\'e approximant on the positive real axis. As a matter of fact one needs to 
select time-steps that are far enough from this spurious pole of the Pad\'e approximant to  avoid catastrophic loss of precision. 

To determine the stepsize in a suitable way, we consider the residual 
\begin{equation}
\mathcal{R}[\tilde{\bf X}] \equiv \left|\frac{d{\bf \tilde{\bf X}}}{dt} - F(t,  {\bf \tilde{X}}(t)) \right| \geq 0
\end{equation}
obtained by substituting the approximate solution $\tilde{\bf X}(t) \equiv  \mathcal{P}\left[{\bf X}(t)\right]_{[m,n]}$ inside eq.~(\ref{eq:ode}). Since the residual vanishes at $t=t_0$,
$\mathcal{R}[\tilde{\bf X}]_{t=t_0}=0$, we expect it to grow in size at later times $t > t_0$, unless 
$\tilde{\bf X}$ is a solution of eq.~(\ref{eq:ode}) (in this case it vanishes identically). In other words, the residual grows monotonically in a sufficiently small region to the right of $t_0$ and we can constrain
the loss of precision of the numerical solution for $t_0 < t < t_0 +\tau_0$, by choosing $\tau_0$ in a suitable way.

This is done by introducing a real parameter $\delta$, with $0 < \delta \ll 1$,  such that $\mathcal{R}[\tilde{\bf X}] \leq \delta$ for $t_0 \leq t \leq t_0+dt_0$~\footnote{If  the equation $\mathcal{R}[\tilde{\bf X}] = \delta$  has multiple real roots we select the root closest to $t_0$, on the right.}.
The value of the time step $dt_0$ found in this way ensures that the error made in approximating the exact solution 
with ${\bf \tilde{X}}$ is bounded at all times $t_0 \leq t \leq t_0+dt_0$. 

The process can now be iterated: moving at the time $t_1=t_0+dt_0$ we can repeat the previous steps by considering eq.~(\ref{eq:ode}) with initial conditions ${\bf X}(t_1)= {\bf \tilde{X} }(t_1)$ and find a new time step and consequently new initial conditions at a time $t_2= t_1+dt_1$.

At each stepsize there is the freedom of changing the order of the Pad\'e approximant used in the calculation, so that we can regard our method as a variable stepsize variable order Pad\'e-Taylor  (VSVO-PT) method.

In the next section we illustrate the method by applying it to the van der Pol equation.

\section{The van der Pol oscillator}
\label{sec:vdP}

The van der Pol equation~\cite{VanderPol28}
\begin{equation}
\ddot{x}(t) + x(t) = \mu \dot{x}(t) \left(1- x^2(t) \right)  \hspace{1cm}, \hspace{1cm} \mu\geq 0
\label{eq_vdp}
\end{equation}
is an example of nonlinear ordinary differential equation that supports a limit cycle: for $\mu >0$, regardless of the 
initial conditions  (with the exclusion of $x(0)=\dot{x}(0)=0$), the  system evolves in time towards a limit cycle  
with fixed period and amplitude (both functions of the parameter $\mu$).
The trajectory that the solution describes in phase space spirals in or out (depending on the initial conditions) towards a close orbit that for $0 <\mu \ll 1$ closely resembles that of a harmonic oscillator while for $\mu \gg 1$  results in a highly deformed loop (reflecting the presence of two very different -- slow and fast -- regimes).

\begin{figure}
\begin{center}
\includegraphics[width=10cm]{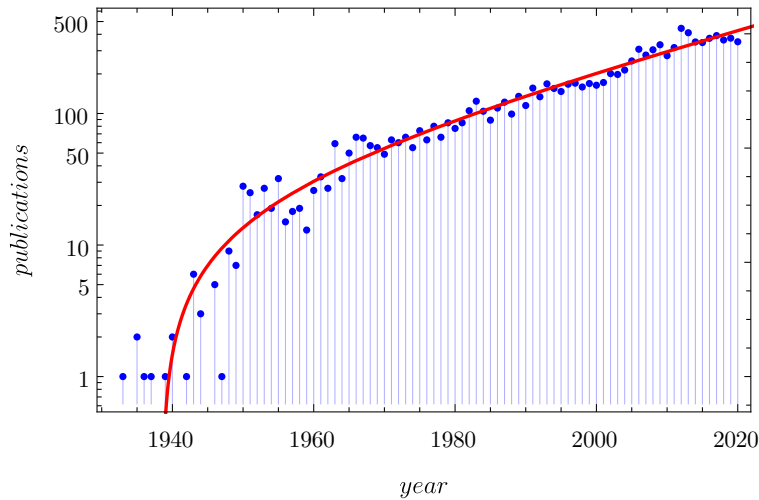}
\caption{Publications per year found with a search of the keyword "van der Pol equation" in google scholar. The solid line is the fit
$N(t) =-28.17+6.31 \times e^{0.034 t-64.66}$.
 }
\label{Fig_google}
\end{center}
\end{figure}

The van der Pol equation constitutes the perfect testing ground for our method, because it represents a challenging 
problem, which has attracted (and continues to attract) much interest in the literature. This interest is reflected in Fig.~\ref{Fig_google} where we report the number of publications per year found in google scholar under the keyword "van der Pol equation". The recent paper by Ginoux and Letellier provides an historical account on the origins of the van der Pol equation~\cite{Ginoux12}.

Following \cite{Amore18} we classify the strategies for studying the limit cycle of the vdP equation into three categories:
\begin{itemize}
\item  numerical studies~\cite{Krogdahl60, Urabe60, Ponzo65b, Zonneveld66, Clenshaw66,Greenspan72}, where the eq. (\ref{eq_vdp}) is solved numerically for some finite value of $\mu$ (apparently the largest value of $\mu$ for which the period and amplitude have been calculated is  $\mu=200$ in ref.~\cite{Ponzo65b}); the accuracy of the results progressively worsens as $\mu$ gets larger~\footnote{In this category we only consider papers that concern the numerical calculation of the period and the amplitude of the vdP oscillator: the vdP equation has been considered in a much larger set of publications, particularly in the stiff regime, as an ideal testing ground for numerical techniques(see for example ref.~\cite{Wanner96,Anne04,Eriksson04,Fazio08,Boom18}).};
\item asymptotic studies~\cite{Haag43, Haag43b, Haag44, Dor47, Dor52, Urabe63, Ponzo65, Ponzo65b, OMalley68,MacGillivray83a,MacGillivray83b}, where the behavior of the solution (particularly the period and the amplitude of the limit cycle) is established for $\mu \rightarrow \infty$; it is  remarkable that as early as in 1944 Haag~\cite{Haag44} and then in 1947 Dorodnitsyn~\cite{Dor47} were able to approximate the asymptotic behavior of the 
period and the amplitude of the limit cycle of the van der Pol oscillator. For a discussion of the  work of Dorodnitsyn, see ref.~\cite{Kerimov11}, written to celebrate the centenary of the birth of A.A. Dorodnitsyn.
\item perturbative studies~\cite{Deprit67,Deprit68,Deprit79,Andersen82,Dadfar84,Buonomo98,Amore04,Suetin12,Amore18}, where the behavior of the solution is established for $\mu \rightarrow 0$ (the perturbative series is a power series in $\mu$ with finite radius of convergence);
\end{itemize}

In the first category we find the works of Krogdahl~\cite{Krogdahl60},  Urabe~\cite{Urabe60}, Zonnenfeld~\cite{Zonneveld66}, Clenshaw~\cite{Clenshaw66} and Greenspan~\cite{Greenspan72} (all published between 1960 and 1972!).
Note that the second method used by Greenspan in \cite{Greenspan72} to solve numerically eq.~(\ref{eq_vdp}) is  a Taylor method of order $8$. 

Later works on the van der Pol oscillator are mainly interested in the study of the perturbation series for the solution, which is most efficiently obtained by applying the Lindstedt-Poincar\'e method.  
The numerical results of refs.~\cite{Krogdahl60,Urabe60,Zonneveld66,Clenshaw66,Greenspan72} are used in those works to validate the accuracy of the resummation schemes adopted (see for instance Table 4 of \cite{Andersen82}, Table 2 of \cite{Dadfar84} and Table 2 of \cite{Buonomo98}). Strictly speaking, the use of the perturbation series for the period (amplitude) of the oscillator is limited to the region $\mu \leq \mu_{T}$ ($\mu \leq \mu_{A}$), where $\mu_{T}$ is the radius  of convergence of the series~\footnote{Remarkably, ref.~\cite{Amore18} contains the most precise determination of $\mu_T$ and $\mu_A$, together with a strong numerical evidence that $\mu_T^2 = \mu_A^2 \approx 3.420 187 909 357 \dots$ (see ref.~\cite{Amore18} for the value of $\mu$ accurate to $100$ digits).}. 
For $\mu > \mu_{T,A}$, different techniques can be applied to improve the convergence of the series, with some success (see \cite{Andersen82, Buonomo98, Amore18}). In this regime however the numerical results are far more precise than the results based on perturbation theory, which can explain only in part the lack of interest in deriving far more precise numerical results. In our view another reason for the scarcity of numerical results is the difficulty in obtaining accurate numerical solutions of eq.~(\ref{eq_vdp}) in the stiff regime ($\mu \gg 1$).

We will now apply to the van der Pol equation the procedures explained in the previous section for the general case.

We start at $t=t_0$, where the initial conditions $x(t_0)=x_0$ and $\dot{x}(t_0) = x'_0$ are provided and consider 
the power series solution
\begin{equation}
x(t) = \sum_{k=0}^\infty c_k (t-t_0)^k
\label{eq:power_series}
\end{equation}
with $c_0 = x(t_0)$ and $c_1=\dot{x}(t_0)$. The coefficients $c_{k}$ with $k\geq 2$ can be obtained in terms 
of the nonlinear recurrence relation
\begin{equation}
\begin{split}
c_{k+2} &= \frac{-c_k + \mu (k+1) c_{k+1} -   \mu \sum_{k_1=1}^{k+1} \sum_{k_2=0}^{k-k_1+1} k_1 \ c_{k_1}\ c_{k_2}\ c_{k-k_1-k_2+1}}{(k+1)(k+2)} 
\end{split}
\label{eq:recurrence}
\end{equation}

\begin{figure}
\begin{center}
\includegraphics[width=8cm]{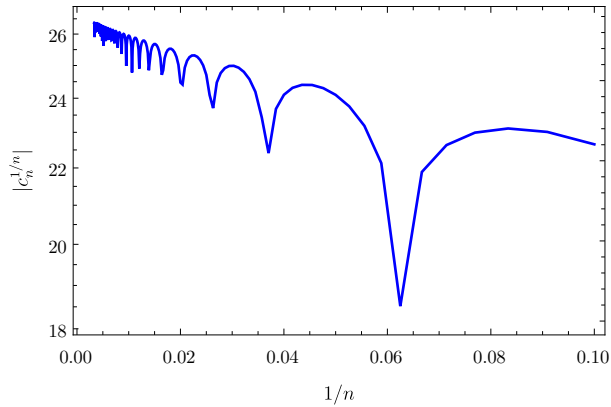}\hspace{0.2cm}
\caption{Approximate radius of convergence of the power series obtained from eq.~(\ref{eq:recurrence}), for $\mu =100$, $x_0=2$ and $x'_0=0$. We plot $1/R \approx |c_n^{1/n}|$ as a function of the order $n$ of the Taylor coefficient (root test).
A similar estimate can be obtained using the Pad\'e approximants. }
\label{Fig_series}
\end{center}
\end{figure}

This equation allows one to calculate exactly a large number of coefficients of the power series and thus estimate its 
radius of convergence. In Fig.~\ref{Fig_series} we estimate the radius of convergence of the power series
(\ref{eq:power_series}) for $\mu =100$, $x_0=2$ and $x'_0=0$, which turns out to be $\approx 1/26$. 

\begin{figure}
\begin{center}
\includegraphics[width=8cm]{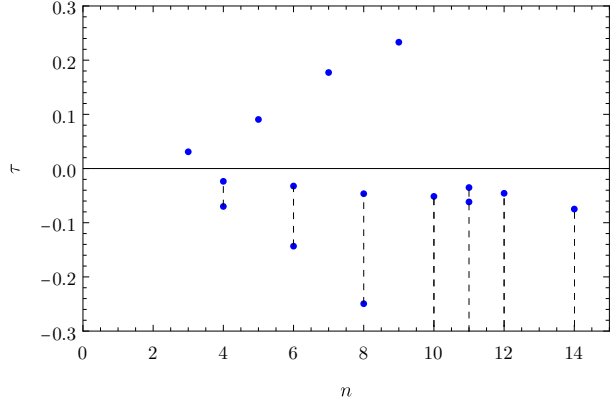}\\		
\caption{Real poles of the Pad\'e approximant of order $[n,n]$ for $\mu =100$, $x_0=2$ and $x'_0=0$. Multiple real roots of the same order are 
joined by a dashed line.}
\label{Fig_series_3}
\end{center}
\end{figure}

In Fig.~\ref{Fig_series_3} we plot the real poles of the Pad\'e approximant of order $[n,n]$ for $\mu =100$, $x_0=2$ and $x'_0=0$: we see that 
Pad\'e of order $3$,$5$,$7$ and $9$ have a positive real root. For example for $n=3$ the root is approximately
$\tau \approx 0.03091366759$, which constitutes an upper limit to the stepsize. Similar behaviors are also observed for smaller values of $\mu$, but in that case the real positive roots are much larger (thus allowing to span a much greater interval of time).

\begin{figure}
\begin{center}
\includegraphics[width=8cm]{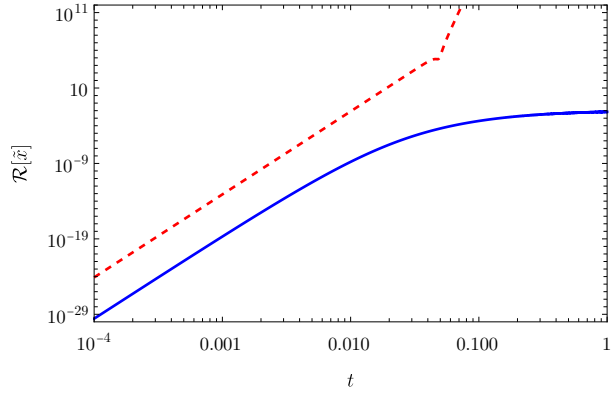}\\		
\caption{Residual obtained using the first $13$ Taylor coefficients for $\mu =100$, $x_0=2$ and $x'_0=0$. The solid line is the result obtained using diagonal
Pad\'e approximants while the dashed line uses the polynomial of order $12$.}
\label{Fig_series_4}
\end{center}
\end{figure}

In Fig.~\ref{Fig_series_4} we plot the residual for $\mu =100$, $x_0=2$ and $x'_0=0$, again using only the first $13$ coefficients: the residual obtained using  a diagonal Pad\'e approximant is about $10^6$ smaller than the corresponding residual built on the 
Taylor polynomial with the same number of coefficients. 
To get an idea, in this case the region for which $\mathcal{R} < 10^{-30}$ is about $3$ times larger when the Pad\'e approximants are used.

There is an important advantage in using our method (or the Taylor method) for the calculation of the 
period and amplitude of the limit cycle (or more in general of any periodic solution): although
these methods work by performing discrete time steps, $t_0 \rightarrow t_1 \rightarrow t_2 \rightarrow \dots$,
the solution is also known at any arbitrary time between any two consecutive times steps, in terms of 
a Pad\'e approximant (in our case) or of a Taylor polynomial (for the Taylor method). In our method, when 
moving from time $t_i$ to time $t_{i+1}$, we expect that the maximum error occurs at $t=t_{i+1}$, where the residual takes the maximum value $\delta$. In this way we can estimate within the same accuracy both the location
of a maximum or a minimum and the value of the solution at this point.

The situation is be completely different using the Runge-Kutta method: for example, Odani~\cite{Odani00}
has estimated the amplitude of the van der Pol oscillator for a limited set of values of the coupling using the Runge-Kutta method with a stepsize of $2^{-20}$! Even so the results reported by Odani contain only $5$ decimals. 
Notice that refs.~\cite{Odani00, Lijun15, Turner15, Cao17, Ignatev17}  deal with the problem of determining the maximum amplitude of the limit cycle.

In our case the procedure to estimate the period and the amplitude of the oscillator is straightforward: we pick the initial conditions (typically we choose $x_0=2$ and $x'_0=0$) and iterate our method until a change in the sign of the velocity is detected; at this point we know that a minimum of the solution can be found between the two last time steps and we can look for the zero of the derivative of the Pad\'e approximant contained in this region. 
The value of the solution at this point (changed of sign) is then used as the new $x_0$ in the initial conditions (we set the time to zero at this point) and the whole procedure is repeated. The time elapsed from $t=0$ to the instant where the derivative changes sign is essentially the half--period. This process is iterated until the amplitude and the period have converged to a sufficiently large number of digits (alternatively the same process can be carried out by looking for two changes of sign in the derivative, which in turn would provide directly the period).

Importantly, we have the nice bonus that the number of iterations that are needed to reach the desired level of accuracy is much smaller in the stiff regime, where the transient behavior of the solution decays more rapidly with time
(the number of time steps however increases, both because of the larger period, which grows with $\mu$, and because of the need for smaller steps in the regions where the derivative is large).

The accuracy that we can hope to reach for the period and the amplitude is essentially determined by 
$\delta$ (provided that the number of digits used in the calculation is sufficiently large): in this way, one
can expect to improve the numerical estimates by using a sufficiently small $\delta$ and working with large enough digits. In the next section we provide an independent check of this claim can be done by comparing the purely numerical results with the analytical results both for $\mu \ll 1$ (perturbative) and for $\mu \gg 1$ (asymptotic -- stiff).

\subsection{Numerical results}
\label{sec:num}

We have obtained numerical results for the period and the amplitude of the limit cycle of the vdP oscillator for a series of values of $\mu$, starting at $\mu=1$ and going up to $\mu=500$ (well beyond the range of values studied in the literature). 
In the tables \ref{table_T_100_vdP}  and \ref{table_A_100_vdP} we report the period and the amplitude of the limit cycle, obtained by working with the Pad\'e-Taylor method, for $\mu = 100$ (stiff), with different values of $\delta$. 
From these tables we see that the results appear to converge quite fast (we expect that the best results in the table, corresponding to $\delta=10^{-100}$, have more than $100$ digits correct).

In separate tables we have reported the values of the period (tables \ref{table_T_vdP}, \ref{table_T_vdP2}, \ref{table_T_vdP3}), 
of the amplitude (tables \ref{table_A_vdP}, \ref{table_A_vdP2}, \ref{table_A_vdP3}), of  the maximum velocity 
(tables \ref{table_vmax_vdP}, \ref{table_vmax_vdP2},\ref{table_vmax_vdP3}), of the fast component of the period~\footnote{By fast component of the period we mean the sum of the times it takes to the oscillator to go from $0$ to a minimum, and from
	$0$ to a maximum, within the limit cycle. For $\mu  \rightarrow \infty$ this time becomes increasingly short, whereas the total period is growing linearly with $\mu$.}
(tables \ref{table_tfast_vdP1}, \ref{table_tfast_vdP2}, \ref{table_tfast_vdP3}) and of $-\lambda_2/\mu$, $\lambda_2$ being the Lyapunov exponent (tables \ref{table_Lyap_vdP1}, \ref{table_Lyap_vdP2}, \ref{table_Lyap_vdP3}) . 

Although Tables \ref{table_T_100_vdP} and \ref{table_A_100_vdP} provide an indication of the accuracy of our results, it is possible to perform an independent check by comparing these numerical results with the perturbative results obtained with the Lindstedt-Poincar\'e method (in ref.~\cite{Amore18} we have worked out analytically the LP to order $308$ and numerically to order $859$ -- in this case the coefficients are calculated with $1000$ digits). 

This comparison can be done directly only for $\mu < \mu_{T,A} \approx 1.85$ (radius of convergence of the perturbative series):
for $\mu=1$ the sum of the first $308$ terms of the LP series for the period and the amplitude are in excellent agreement 
with the numerical results obtained with the PTm:
\begin{equation}
T^{({\rm num})} - T^{(LP)} \approx  3.57 \times 10^{-87}
\end{equation}
and
\begin{equation}
A^{({\rm num})} - A^{(LP)} \approx  1.97 \times 10^{-87}
\end{equation}

The perturbative results, however, can be extended beyond the radius of convergence of the series, by using a Pad\'e approximant 
(we opted for a diagonal Pad\'e $[150,150]$). 
For $\mu= 1$, the agreement between the Pad\'e approximant and the numerical results is now extended to more than $100$ decimals:
\begin{equation}
T^{({\rm num})} - T^{(LP-Pade)} \approx  -8.2 \times 10^{-104} 
\end{equation}
and
\begin{equation}
A^{({\rm num})} - A^{(LP-Pade)} \approx  -5.17 \times 10^{-107} 
\end{equation}

Even for $\mu > \mu_{T,A}$, where the LP series does not converge, there is a remarkable agreement between 
the numerical results and those obtained using the Pad\'e approximants, for $\mu \leq 10$. 

\begin{figure}
\begin{center}
\includegraphics[width=12cm]{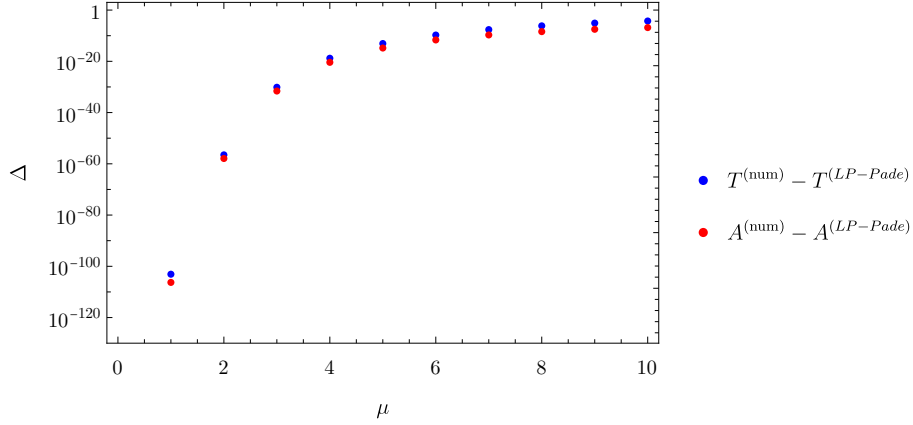}\\		
\caption{$\Delta_T = | T^{({\rm num})}- T^{(LP-Pade)} |$ and $\Delta_A = | A^{({\rm num})}- A^{(LP-Pade)} |$ as functions of $\mu$. Diagonal Pad\'e approximants of order $[150,150]$ are used.}
\label{Fig_error_TA}
\end{center}
\end{figure}

\begin{figure}
\begin{center}
\includegraphics[width=9cm]{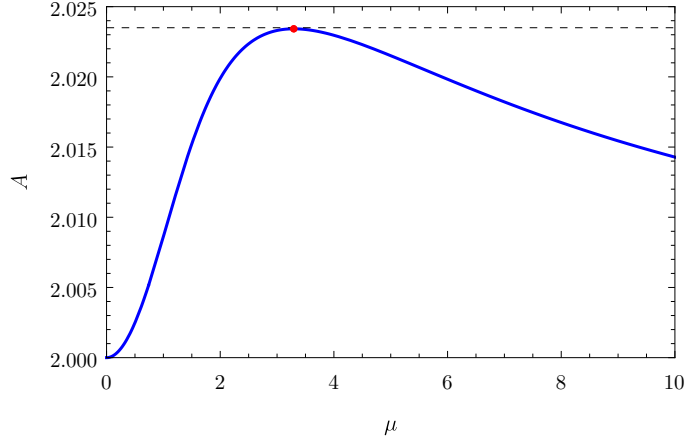}\\		
\caption{Amplitude of the limit cycle of the van der Pol oscillator obtained from the Pad\'e approximant of order $[150,150]$ of the perturbative (LP) expansion. The horizontal dashed line corresponds to the upper bound conjectured by 
Odani~\cite{Odani00}, $A^{\rm (Odani)} = 2.0235$. The red dot is the maximum value taken by the Pad\'e approximant,  $A^{(max)}=2.0234222556061133094$ for $\mu = 3.2940126635728034197$.}
\label{Fig_Max_Ampl_vdp}
\end{center}
\end{figure}

This comparison is performed in Fig.~\ref{Fig_error_TA}, where we appreciate that, for $\mu \leq 4$, the results agree to more than $20$ decimals. This observation justifies using the Pad\'e approximant for the amplitude to estimate quite precisely the maximum amplitude of limit cycle of the van der Pol equation, that has been studied in a series of papers, ref.~\cite{Odani00, Lijun15, Turner15, Cao17, Ignatev17}, but never calculated with high precision. In Fig.~\ref{Fig_Max_Ampl_vdp} we plot the Pad\'e approximant for the amplitude (solid line) and compare it with the upper bound conjectured by Odani ~\cite{Odani00}. 
Our estimated value for the maximum amplitude is $A^{(max)}=2.0234222556061133094$ for $\mu = 3.2940126635728034197$ (which complies with Odani's conjectured upper bound $2.0235$).

\begin{figure}
\begin{center}
\includegraphics[width=11cm]{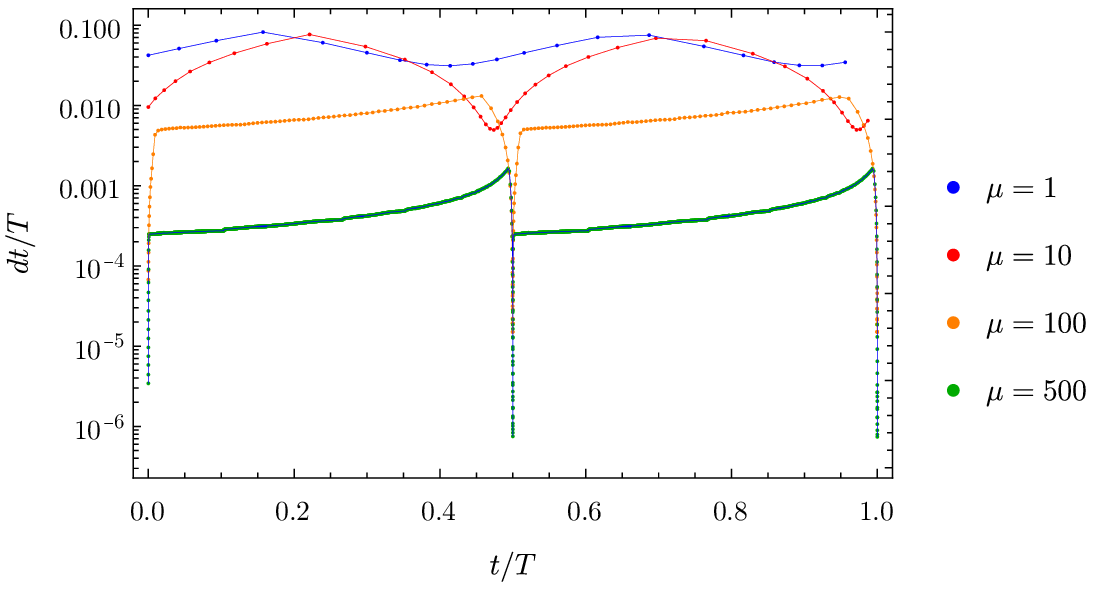}\\		
\caption{Stepsize of the Pad\'e-Taylor method with $\delta=10^{-100}$, using $201$ coefficients, for $\mu=1, 10, 100, 500$.}
\label{Fig_step_vdp}
\end{center}
\end{figure}

\begin{figure}
	\begin{center}
		\includegraphics[width=9cm]{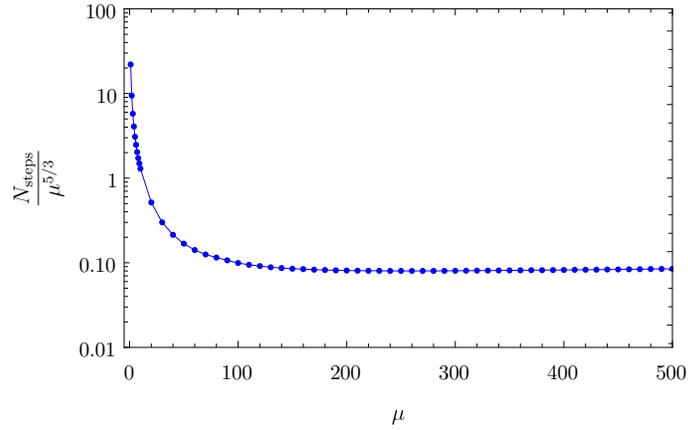}\\		
		\caption{Number of steps used by the Pad\'e-Taylor method with $\delta=10^{-100}$, and $201$ coefficients, as function of $\mu$. }
		\label{Fig_nstep_vdp}
	\end{center}
\end{figure}

\begin{figure}
\begin{center}
\includegraphics[width=11cm]{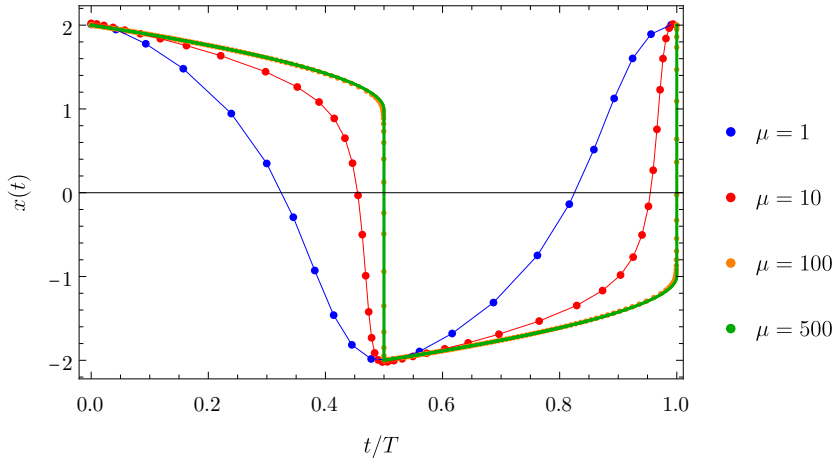}\\		
\caption{Solutions to the vdP equation  for $\mu=1, 10, 100, 500$ using  the Pad\'e-Taylor method with $\delta=10^{-100}$, and $201$ coefficients. }
\label{Fig_sol_vdp}
\end{center}
\end{figure}

In Fig.~\ref{Fig_step_vdp} we compare the stepsize of the PTm (normalized to the period) for different values of $\mu$; observe that the number of steps required to cover the whole period appears to scale as $\propto \mu^{5/3}$ (see Fig.~\ref{Fig_nstep_vdp}). The corresponding solutions are displayed in Fig.~\ref{Fig_sol_vdp}.

The reader may be skeptical on the necessity of calculating the period, the amplitude and more in general the limit cycle with such large precision as we claim having reached, but we hope that our next application will convince the most critical reader. 

\subsection{Extrapolation}
\label{sec:extra}

In this section we adapt the discussion of Ref.~\cite{Amore16}, where the Richardson extrapolation was applied to the finite difference calculation of the Laplacian eigenvalues on complicated non-tensor domains in two dimensions.

In the previous section we have obtained very precise numerical results  for different quantities (period, amplitude, maximum velocity, Lyapunov exponent, etc) related to the limit cycle of the van der Pol equation for different values of $\mu$.
Let $\mathcal{O}$ be such quantity and consider a sequence of values $\mathcal{O}_1$, $\mathcal{O}_2$, \dots, corresponding to different values  of the parameter, $\mu_1$, $\mu_2$, \dots.

For $\mu \gg 1$, the behavior of $\mathcal{O}(\mu)$ may be approximated as
\begin{equation}
\mathcal{O}(\mu) \approx \sum_{k=0}^\infty c_k f_k(\mu) = c_0 f_0(\mu) + c_1 f_1(\mu) + \dots 
\end{equation}
where $f_0(\mu) \gg  f_1(\mu) \gg f_2(\mu) \gg \dots$, for $\mu \rightarrow \infty$.  For the case $\mathcal{O}(\mu) = T(\mu)$, for example, several terms of the asymptotic formula has been derived 
in \cite{Haag43,Haag43b,Haag44,Dor47,Dor52,Urabe63,Ponzo65} and correspond to choosing
\begin{equation}
\begin{split}
f_0(\mu) &= \mu \\ 
f_1(\mu) &= \mu^{-1/3} \\ 
f_2(\mu) &= \frac{\log\mu}{\mu} \\
f_3(\mu) &= \frac{1}{\mu} 
\label{eq_extrapolation_f}
\end{split}
\end{equation}

If we dispose of $N$ values of $\mathcal{O}$, for different values of $\mu$, and restrict our calculation to the first $N$ functions $f_i(\mu)$, we can write the system of equations 
\begin{equation}
\begin{split}
\mathcal{O}_1 &=   c_0 f_0(\mu_1) + c_1 f_1(\mu_1) + \dots  + c_{N-1} f_{N-1}(\mu_1) \\
\mathcal{O}_2 &=   c_0 f_0(\mu_2) + c_1 f_1(\mu_2) + \dots  + c_{N-1} f_{N-1}(\mu_2) \\
&\dots \\
\mathcal{O}_N &=   c_0 f_0(\mu_N) + c_1 f_1(\mu_N) + \dots  + c_{N-1} f_{N-1}(\mu_N) \\
\end{split}
\label{eq_system}
\end{equation}
which can be cast in matrix form as
\begin{equation}
\mathbf{R} \left(
\begin{array}{c}
c_0 \\
c_1 \\
\dots \\
c_{N-1} \\
\end{array}
\right) = \left(
\begin{array}{c}
\mathcal{O}_1  \\
\mathcal{O}_2 \\
	\dots \\
\mathcal{O}_{N-1}\\
\end{array}
\right) 
\end{equation}
where 
\begin{equation}
\mathbf{R} \equiv \left( 
\begin{array}{cccc}
f_0(\mu_1) & f_1(\mu_1) & \dots & f_{N-1}(\mu_1) \\
f_0(\mu_2) & f_1(\mu_2) & \dots & f_{N-1}(\mu_2) \\
\dots  & \dots  & \dots  & \dots  \\
f_0(\mu_N) & f_1(\mu_N) & \dots & f_{N-1}(\mu_N) \\
\end{array}
\right)
\end{equation}

The solution to eq.~(\ref{eq_system}) can be obtained either by using the inverse matrix $\mathbf{R}^{-1}$
\begin{equation}
 \left(
\begin{array}{c}
	c_0 \\
	c_1 \\
	\dots \\
	c_{N-1} \\
\end{array}
\right) = \mathbf{R}^{-1} \left(
\begin{array}{c}
	\mathcal{O}_1  \\
	\mathcal{O}_2 \\
	\dots \\
	\mathcal{O}_{N-1}\\
\end{array}
\right) 
\end{equation}
or by using Cramer's  rule.

A further improvement can be obtained by performing a Richardson-Pad\'e extrapolation, as done in ref.~\cite{Amore16}; 
in this case we  assume 
\begin{equation}
\mathcal{O}(\mu) \approx \frac{\sum_{k=0}^N c_k f_k(\mu)}{1+\sum_{k=1}^M d_k g_k(\mu)}
\end{equation}
where $\lim_{\mu \rightarrow \infty } g_k(\mu) =0$ (a convenient choice is $g_k(\mu) = \frac{1}{\mu^k}$, $k=1,2,\dots$).
By using a rational extrapolation, we may be able to describe more precisely also the behavior of the observable for smaller values of $\mu$.

Using $N+M+1$ numerical results one can write the system of equations
\begin{equation}
\begin{split}
\mathcal{O}_1 &=  c_0 f_0(\mu_1) + c_1 f_1(\mu_1) + \dots  + c_{N-1} f_{N-1}(\mu_1) \\
&- \left[ d_1 g_1(\mu_1) + \dots + d_M g_M(\mu_1)\right] \mathcal{O}_1 \\
\mathcal{O}_2 &=  c_0 f_0(\mu_2) + c_1 f_1(\mu_2) + \dots  + c_{N-1} f_{N-1}(\mu_2) \\
&- \left[ d_1 g_1(\mu_2) + \dots + d_M g_M(\mu_2)\right] \mathcal{O}_1 \\
&\dots \\
\mathcal{O}_{N+M+1} &=  c_0 f_0(\mu_{N+M+1} ) + \dots  + c_{N-1} f_{N-1}(\mu_{N+M+1} ) \\
&- \left[ d_1 g_1(\mu_{N+M+1} ) + \dots + d_M g_M(\mu_{N+M+1} )\right] \mathcal{O}_1 \\
\end{split}
\end{equation}
which can be cast in matrix form as
\begin{equation}
\tilde{\mathbf{R}}  \left(
\begin{array}{c}
	c_0 \\
	\dots \\
	c_{N} \\
	d_1 \\
	\dots \\
	d_{M} \\
\end{array}
\right) = \left(
\begin{array}{c}
	\mathcal{O}_1  \\
	\mathcal{O}_2 \\
	\dots \\
	\mathcal{O}_{N+M+1}\\
\end{array}
\right) 
\end{equation}
where
\begin{equation}
\begin{split}
\tilde{\mathbf{R}} \equiv \left(
\begin{array}{cccccc}
f_0(\mu_1) & \dots & f_{N-}(\mu_1) & - g_1(\mu_1) \mathcal{O}_1 & \dots & - g_M(\mu_1) \mathcal{O}_1  \\
f_0(\mu_2) & \dots & f_{N-}(\mu_2) & - g_1(\mu_2) \mathcal{O}_1 & \dots & - g_M(\mu_2) \mathcal{O}_2  \\
\dots &  \dots &  \dots &  \dots & \dots & \dots \\         
f_0(\mu_{N+M+1}) & \dots & f_{N-}(\mu_{N+M+1}) & - g_1(\mu_{N+M+1}) \mathcal{O}_{N+M+1} & \dots & - g_M(\mu_{N+M+1}) \mathcal{O}_{N+M+1}  \\
\end{array}
\right)
\end{split}
\end{equation}
Once again we can solve these equations either by using the inverse of $\tilde{\mathbf{R}}$ or by applying Cramer's rule.

Of course the implementation of the Richardson and Richardson-Pad\'e extrapolation requires to identify as many $f_i(\mu)$ as possible, which may be  a difficult task in general. Some of these
functions may be provided by previous asymptotic calculations, as for the case of the period and the amplitude of the limit cycle, but the remaining functions need to be "guessed". 
In an empirical approach one may select functions that lead to a faster convergence of the extrapolated results and discard functions whose coefficients are abnormally small.

Determining  these functions rigorously is most likely a very difficult task: for the case of the Dirichlet eigenvalues of the Laplacian on a L-shaped membrane, for which the first correction 
to the lowest eigenvalue has been proved to behave as $h^{4/3}$ ($h$ is the grid spacing),  the next few non-integer exponents have not been derived rigorously and, quoting 
Kuttler and Sigillito \cite{Kuttler84}, “the exact form of the ﬁrst several terms in the asymptotic formula for
 speciﬁc regions where no boundary interpolation is required is a nice problem at about the level of a doctoral thesis”. The fact that this problem is still open can be taken as an indication of its difficulty.

The extrapolation carried out in \cite{Amore16}, where we tried to guess these exponents, resulted in obtaining very precisely the lowest eigenvalues of the L-shaped
membrane (later confimed by the results obtained in ref.~\cite{Jones17} using a different method) and the most precise estimate to date of the lowest eigenvalues of the pair of isospectral domains
of ref.~\cite{Gordon92} (see also Refs.~\cite{Amore19,Amore19b} for further example of applications of the Richardson and Richardson-Pad\'e extrapolations).

In Tables \ref{table_tfast_vdP1}, \ref{table_tfast_vdP2} and \ref{table_tfast_vdP3} we report the precise numerical values of the times 
$T_{\rm fast}$ for different values of $\mu$.
 
We perform a Richardson extrapolation using the functions 
\begin{equation}
\begin{split}
f_0(\mu) = \frac{\log \mu}{\mu} \hspace{0.3cm} , \hspace{0.3cm} f_1(\mu) &=  \frac{1}{\mu}  \hspace{0.3cm} , \hspace{0.3cm} f_2(\mu) = \frac{1}{\mu^2} \hspace{1cm }\dots \\
\end{split}
\end{equation} 
and use the last $50$ numerical values to estimate the leading behavior of $T_{\rm fast}(\mu)$ for $\mu \gg 1$.

Our analysis shows that 
\begin{equation}
	T_{\rm fast}(\mu) \approx 1.33333333333333333334 \ \frac{\log\mu}{\mu} + \frac{3.5305579106695527878}{\mu}  + \dots
\end{equation}
where one can observe that $1.33333333333333333334-4/3 \approx 6.6 \times 10^{-21}$ and $3.5305579106695527878 - \left(\frac{4}{3}+\log 9\right) \approx 7.2 \times 10^{-17}$.

We then have
\begin{equation}
T_{\rm fast}(\mu) \approx \frac{4}{3} \ \frac{\log\mu}{\mu} + \left(\frac{4}{3}+\log 9\right) \frac{1}{\mu}  + \dots
\end{equation}

The slow component of the period, $T_{\rm slow}$ is obtained from the condition $T_{\rm slow} = T - T_{\rm fast}$. In this case
the Richardson extrapolation provides 
\begin{equation}
\begin{split}
T_{\rm slow}(\mu) &\approx  1.6137056388801093811656 \ \mu  +\frac{7.0143222313792851710947}{\mu^{1/3}} \\
&- 1.99999999955 \frac{\log (\mu )}{\mu } -\frac{4.853823019}{\mu}  + \dots
\end{split}
\end{equation}

The accuracy of the first $3$ numerical coefficients allows to guess their exact values
\begin{equation}
\begin{split}
T_{\rm slow}(\mu) \approx (3-\log 4)  \mu - \frac{3 \alpha}{\mu^{1/3}}  - 2 \frac{\log \mu}{\mu}-\frac{4.853823019}{\mu}  + \dots
\end{split}
\end{equation}
where $\alpha$ is the first zero of the Airy function ${\rm Ai}(x)$.

We can improve the estimate of the last coefficient by performing the Richardson extrapolation on the data where the contributions of the three
first terms is taken out: in this case we obtain the slightly better value $-4.85382301173$.

By combining the formulas for the fast and slow periods, we obtain the asymptotic formula for the period of the vdP oscillator in the form
\begin{equation}
	\begin{split}
		T(\mu) \approx (3-\log 4)  \mu - \frac{3 \alpha}{\mu^{1/3}}  - \frac{2}{3} \frac{\log \mu}{\mu}-\frac{1.323265101}{\mu}  + \dots
	\end{split}
\end{equation}
which coincides with eq.~(4) of Ref.~\cite{Ponzo65b}, apart from last term that reads $-\frac{1.3246}{\mu}$. The coefficient $-1.3246$ used by Ponzo and Wax had been estimated earlier by Urabe~\cite{Urabe60}. 

It is interesting to observe that the formulas for the period given by Dorodnitcyn~\cite{Dor47,Dor52} and Urabe~\cite{Urabe60} contain a different
coefficient for the term $\log \mu/\mu$, $-22/9$ and $-1/3$ respectively. Ponzo and Wax were able to establish what is now accepted as the correct coefficient $-2/3$ by performing a numerical calculation up to $\mu = 200$ (see Table 1 of ref.~\cite{Ponzo65b}).

\begin{figure}
\begin{center}
\includegraphics[width=8cm]{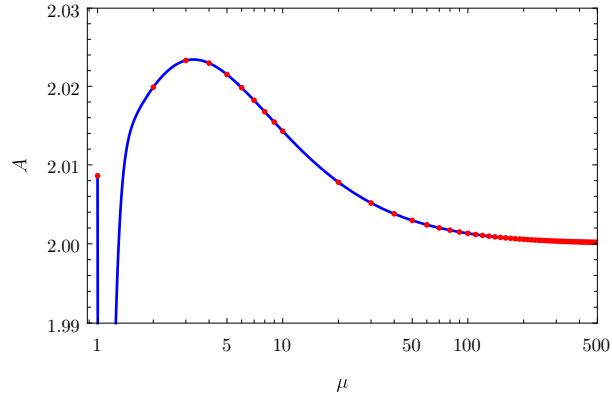}
\caption{Richardson-Pad\'e extrapolation for the amplitude of the limit cycle of the vdP oscillator (solid line). The dots are the numerical values for the amplitude reported in Tables \ref{table_A_vdP}, \ref{table_A_vdP2} and \ref{table_A_vdP3}.}
\label{Fig_Pade_A}
\end{center}
\end{figure}

By performing a Richardson-Pad\'e extrapolation we are able to describe the numerical values for the amplitude in an excellent way, up to $\mu \approx 2$ (see Fig.~\ref{Fig_Pade_A}). In fact the Richardson-Pad\'e extrapolation is so accurate that it reproduces the first $10$ decimals of the value of the maximum amplitude obtained using the Pad\'e approximant of order $[150,150]$ for the Lindstedt-Poincar\'e expansion! Similarly the value of $\mu$ corresponding to this maximum agrees within $8$ decimals with the value previously obtained. 

While the expression for the amplitude obtained with Richardson-Pad\'e is too lengthy to be reported here, it is useful to 
approximate it for $\mu \gg 1$, where it reads:
\begin{equation}
\begin{split}
A &\approx 2.000000000000000000000084 + \frac{0.77936913681989}{\mu^{4/3}} \\
&-\frac{0.5925925919 \log (\mu)}{\mu^2} -\frac{0.8761741716}{\mu^2} \\
&- \frac{0.1988699974}{\mu^{8/3}} +\frac{0.1914226901}{\mu^3}-\frac{0.01149412656 \log (\mu )}{\mu^3} +\dots
\end{split}
\end{equation}

It is easy to guess the exact form for the first three coefficients and obtain
\begin{equation}
\begin{split}
A &\approx 2 - \frac{\alpha}{3 \mu^{4/3}}  - \frac{27}{16} \frac{\log\mu}{\mu^2}
 -\frac{0.8761741716}{\mu^2} \\
&- \frac{0.1988699974}{\mu^{8/3}} +\frac{0.1914226901}{\mu^3}-\frac{0.01149412656 \log (\mu )}{\mu^3} +\dots
\end{split}
\end{equation}
which agrees with the asymptotic formula for the amplitude, while adding few extra terms.

In the case of the maximum of $\dot{x}$ (corresponding to the numerical results in Tables \ref{table_vmax_vdP}, \ref{table_vmax_vdP2} and \ref{table_vmax_vdP3}) using Richardson extrapolation we have found, on purely numerical grounds, the behavior
\begin{equation}
\max\left\{\dot{x}\right\} \approx  \frac{4}{3} \mu - \frac{\alpha}{\mu^{1/3}} + \frac{4}{9} \frac{\log \mu}{\mu} + \dots
\end{equation}
This formula has not been obtained before.

\begin{figure}
\begin{center}
\includegraphics[width=12cm]{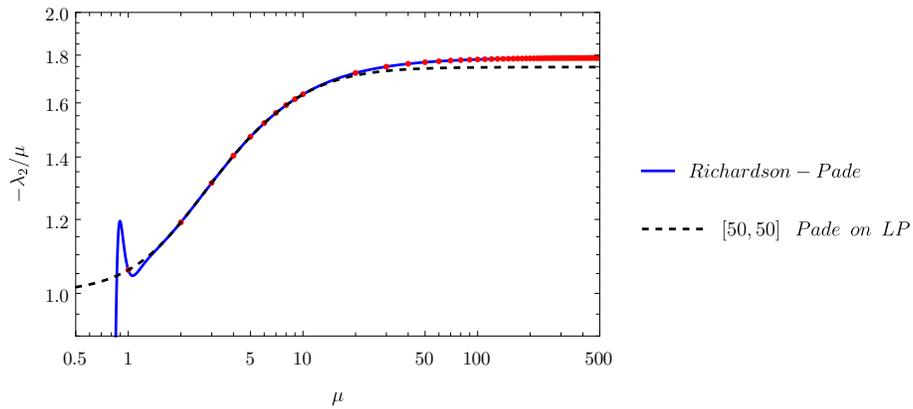}
\caption{$-\lambda_2/\mu$ as a function of $\mu$ ($\lambda_2$ is the Lyapunov exponent). The dots are the precise numerical results obtained with the TPm; the solid blue line is the Richardson-Pad\'e extrapolation of the numerical results, whereas the dashed line is the $[50,50]$ Pad\'e approximant of the Lindstedt-Poincar\'e series.}
\label{Fig_Lyapunov}
\end{center}
\end{figure}

\begin{figure}
\begin{center}
\includegraphics[width=8cm]{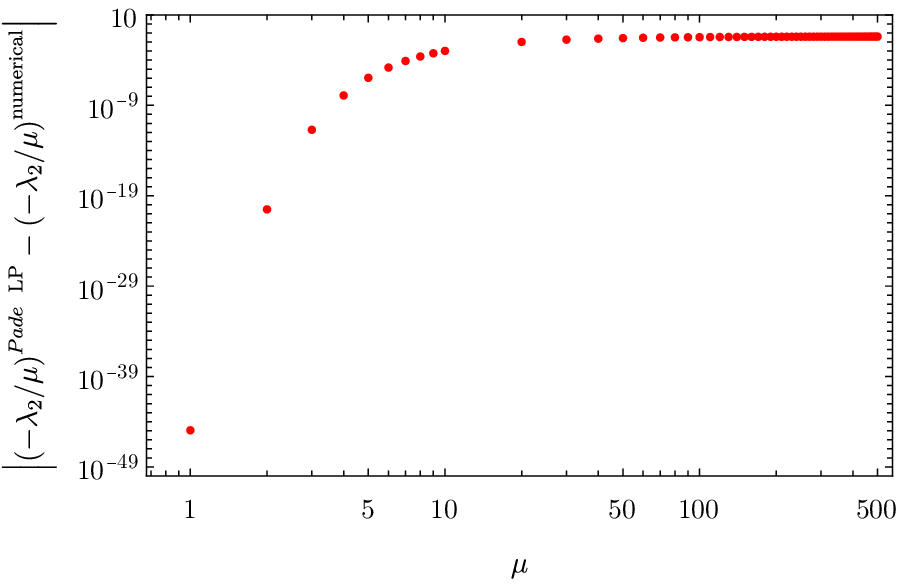}
\caption{$\left|(-\lambda_2/\mu)^{Pade \ {\rm LP}}- (-\lambda_2/\mu)^{{\rm numerical}}\right|$ as a function of $\mu$.}
\label{Fig_Lyapunov_error}
\end{center}
\end{figure}

Finally we consider the Lyapunov exponent, for which the asymptotic behavior has been established in ref.~\cite{Grasman05}.
By performing the Richardson-Pad\'e extrapolation of the numerical results, we have found that, for $\mu \gg 1$, the 
quantity $-\lambda_2/\mu$ behaves approximately as
\begin{equation}
\begin{split}
- \lambda_2/\mu &\approx  1.7886126760534475020333538 - 3.427876368721168125 \mu^{-4/3} \\
&- 0.913585588576346183705582275127144801985703792607697195  \frac{\log (\mu)}{\mu ^2} \\
&- \frac{1.016871272807540550530978993562345482080521280}{\mu^2}  + \dots
\end{split}
\end{equation}

The coefficient of the term $\mu^0$ agrees to $24$ digits with the predicted coefficient of \cite{Grasman05}, $\frac{3+\log (16)}{6-\log (16)}$; the coefficient of the term $\mu^{-4/3}$ agrees to $15$ digits with the predicted coefficient of \cite{Grasman05},
$-\frac{3 (\log (256)-3) \alpha}{2 (\log (4)-3)^2}$. 

Alternatively one can obtain a perturbative expression for $-\lambda_2/\mu$ using the Lindstedt-Poincar\'e method; 
using the results previous obtained in \cite{Amore18} we can write  the first few terms read~\footnote{We have calculated the coefficients up to order $\mu^{100}$, but we do not report them here.}
\begin{equation}
\left. -\frac{\lambda_2}{\mu}\right|_{\rm LP} = 1+\frac{\mu^2}{16}-\frac{11 \mu^4}{4608}-\frac{4859 \mu^6}{5308416}+\frac{12921629 \mu^8}{76441190400}+\frac{22269589099\mu ^{10}}{1100753141760000} + O\left(\mu^{12}\right)
\label{eq_Lyapunov_LP}
\end{equation}

Notice that our coefficient of order $\mu^4$ does not agree with the corresponding coefficient of eq.(14) of \cite{Grasman05}.
In Fig.~\ref{Fig_Lyapunov} we plot the numerical results for $-\lambda_2/\mu$ and compare them with the Pad\'e approximant of order $[50,50]$ built on the perturbative LP series (dashed line)
and the Richardson-Pad\'e extrapolation of the numerical results (solid line).
In Fig.~\ref{Fig_Lyapunov_error} we plot the quantity $\left|(-\lambda_2/\mu)^{Pade \ {\rm LP}}- (-\lambda_2/\mu)^{{\rm numerical}}\right|$. We observe a remarkable agreement between the perturbative Pad\'e approximant and the numerical results for $\mu \leq 5$. The numerical results presented in \cite{Grasman05} appear to have limited accuracy (even for $\mu=1$, the result reported in table I of \cite{Grasman05}, $-\frac{\lambda_2}{\mu} \approx 1.0648$, appears to have only two digits correct).

\section{Conclusions}
\label{sec:concl}

The sentence that stiff ODE cannot be solved by means of explicit methods, such as the Taylor method, is sometimes used as a definition
of stiffness (pag.2 of ref.~\cite{Wanner96}). In particular, refs.~\cite{Jorba05} and \cite{Barrio05}, that describe  explicit Taylor methods, also state this limitation and consider only non--stiff applications.

In this paper we have devised a different explicit method, that uses Pad\'e approximants rather than Taylor series; our method does not suffer from this limitation and in fact it can be used to obtain arbitrarily accurate numerical solutions both  for stiff and 
non--stiff problems. The accuracy of the solutions is limited only by the computing time and by the computer memory, provided that one works with a sufficiently large number of digits. 

To illustrate the virtues and potentialities of our method we have applied it to the van der Pol equation, that is often used as an 
example of stiff ODE. The advantage of considering this equation lies on the fact that there it has been widely studied, both numerically and analytically: in particular, the perturbative Lindstedt-Poincar\'e expansion has been calculated to very high order in our previous paper \cite{Amore18}, while the asymptotic behavior for large coupling has been established with the pioneer work of 
Dorodnitcyn~\cite{Dor47,Dor52} and subsequent improvements by Urabe~\cite{Urabe60} and by Ponzo and Wax~\cite{Ponzo65b}. All the numerical calculations of both period and amplitude of the vdP oscillator are quite old and with a  limited precision. Despite the fact
that the vdP equation is a standard example in many textbooks, we are unaware of recent attempts in improving these numerical results.
With the exception of ref.~\cite{Ponzo65b}, where the numerical experiment has lead to the determination of the correct coefficient
of the term $1/\mu^2$ in the expression for the period, the numerical approximations have only been used to provide a qualitative
validation of the asymptotic behaviors of the period and amplitude.

By disposing with our method of extremely precise numerical results (roughly $100$ digits) we have been able to infer the asymptotic
behaviors of the period and of the amplitude of the vdp oscillator by means of both Richardson and Richardson-Pad\'e extrapolations.
The expressions for the period and the amplitude reproduce the asymptotic formulas of \cite{Dor47,Dor52,Urabe60,Ponzo65b}. 
Similarly our extrapolation of the Lyapunov exponent agrees with the formula of ref.~\cite{Grasman05}.

Additionally, we have been able to  infer
the asymptotic behavior of the portion of the period corresponding to fast (slow) motion, and of the maximum amplitude of the velocity,
which have never  been derived before. It would be nice that our numerical work could stimulate the interest in deriving such behaviors from an asymptotic analysis. 

We plan to apply our method to discuss further examples in future work.

\section*{Acknowledgements}
The research of P.A. was supported by Sistema Nacional de Investigadores (M\'exico). 
The plots in this paper have been plotted using {\rm MaTeX} \cite{szhorvat}. Numerical calculations have been carried out using Mathematica~\cite{wolfram}.

\appendix

\begin{sidewaystable}
	\caption{Period of the limit cycle of the vdP oscillator calculated with the PTM for $\mu=100$ for different values of $\delta$, using Pad\'e approximants of order $[100,100]$. We are working with $400$ digits of accuracy. Underlined digits have converged to the result for $\delta=10^{-100}$.}
	\bigskip
	\label{table_T_100_vdP}
	\begin{tabular}{|c|l|}
		\hline
		$-\log \delta$ &   $T$    \\
		\hline
		10	&   \underline{162.837071092370012132463707167168616680181839}251756941333330629148899313424297715660870141334674 \\
		30	&   \underline{162.837071092370012132463707167168616680181839103855961135766672854022692827637159}152988577327590 \\
		50	&   \underline{162.837071092370012132463707167168616680181839103855961135766672854022692827637159990178783311025} \\
		&   \hspace{0.5cm} \underline{2564}4658408351681621609412185598512187657024 \\
		70	&   \underline{162.837071092370012132463707167168616680181839103855961135766672854022692827637159990178783311025} \\
		&   \hspace{0.5cm} \underline{2564657527763077908305}3709785049344630790442 \\
		90 	&   \underline{162.837071092370012132463707167168616680181839103855961135766672854022692827637159990178783311025} \\
		&   \hspace{0.5cm} \underline{256465752776307790830529289660508539773033}74 \\
		& \\
		\hline \hline
		& \\
		100 &	162.837071092370012132463707167168616680181839103855961135766672854022692827637159990178783311025 \\
		&   \hspace{0.5cm} 25646575277630779083052928966050853977303377 \\
		\hline
		\hline
	\end{tabular}
	\bigskip\bigskip
\end{sidewaystable}

\begin{sidewaystable}
	\caption{Amplitude of the limit cycle of the vdP oscillator calculated with the PTM for $\mu=100$ for different values of $\delta$, using Pad\'e approximants of order $[100,100]$. We are working with $400$ digits of accuracy. Underlined digits have converged to the result for $\delta=10^{-100}$.}
	\bigskip
	\label{table_A_100_vdP}
	\begin{tabular}{|c|l|}
		\hline
		$-\log \delta$ &   $A$    \\
		\hline
		10	&  \underline{2.00131868117722416123741276567482482666792226255}1957901610567138094890079994152788556570113501359 \\
		30	&  \underline{2.00131868117722416123741276567482482666792226255218608294786066364594471854125707587482}92891139 \\
		50	&  \underline{2.001318681177224161237412765674824826667922262552186082947860663645944718541257075874824291065133} \\
		&  \hspace{0.5cm}  \underline{5970990091517}397119495895612305437072973177 \\
		70	&  \underline{2.001318681177224161237412765674824826667922262552186082947860663645944718541257075874824291065133} \\    
		&  \hspace{0.5cm} \underline{5970990091517401667571422742334320536}187643 \\
		90	&  \underline{2.001318681177224161237412765674824826667922262552186082947860663645944718541257075874824291065133} \\
		&  \hspace{0.5cm} \underline{5970990091517401667571422742334320536051373070050188917087}61643 \\
		& \\
		\hline \hline
		& \\
		100	& 2.001318681177224161237412765674824826667922262552186082947860663645944718541257075874824291065133\\
		&  \hspace{0.5cm} 597099009151740166757142274233432053605137307005018891708741188 \\
		\hline
		\hline
	\end{tabular}
	\bigskip\bigskip
\end{sidewaystable}

\begin{sidewaystable}
	\caption{Period of the limit cycle of the vdP oscillator calculated with the PTM with $\delta=10^{-100}$ and using Pad\'e approximants of order $[100,100]$. We are working with $400$ digits of accuracy. }
	\bigskip
	\label{table_T_vdP}
	\begin{tabular}{|c|c|}
		\hline
		$\mu$ &   $T$    \\
		\hline
		1   &  6.663286859323130189699682030482328706812646316883876655114863182080925864845606864868966620053271223 \\
		2   &  7.629874479674841631969395257463658522428587629692901447701015386657008410006341189416420810954316218 \\
		3   &  8.859095499719917151051514310463877906020714102264039268895369779521275441233406049867411814504144379 \\
		4   & 10.20352369099367422324899444326529091932898562514092695707663779128889460634210527904382560843475084  \\
		5   & 11.61223066771957003455506622973852318262711328074405333346140310979401962620255595598026664353518727  \\
		6   & 13.06187474255027210113752236149315139265846873873044360016428664371608166229543648433625630756910465  \\
		7   & 14.53974774448424937257021485659788884491227135001325280022234023529434557878813603167335765513955300  \\
		8   & 16.03817623218046573464954653234692073100521038686081165776010304458797832279349566390537465014875784  \\
		9   & 17.55218413562059696674065075709756465487236738456195028205656607301822261169285330210848834642978791  \\
		10  & 19.07836956693901407043011282581517232694950913194955157265580656843824505331274115773388716007112246  \\
		20  & 34.68232331165268356712000011068043241399560335452075397983964255953220814833151481417843551741048142  \\
		30  & 50.54368648274051207028273864668370468382482785860735065389475279356222926957662902442086062638175210  \\
		40  & 66.50136904280447699869719249433878375984381172820748685457193715914988465513250863494756199855998212  \\
		50  & 82.50833389323078218683336869749987459374059085861862778675385441021240035170950232648047837387499695  \\ 
		60  & 98.54478895887890341182365683019522186958016650440952312516595456004299353306650563194910715150571525  \\
		70  & 114.6006663004441624691712023317540856185451162034576279580190523911294876452269860606383216961761215  \\
		80  & 130.6701966309285473096077258773381797819703833326686760165963074383845409405823897922106445547563429  \\
		90  & 146.7497896543084106513374421250368854262190114879122183612587391565662535580686126367442628577020044  \\
		100 & 162.8370710923700121324637071671686166801818391038559611357666728540226928276371599901787833110252565  \\
		\hline
		\hline
	\end{tabular}
	\bigskip\bigskip
\end{sidewaystable}

\begin{sidewaystable}
	\caption{Period of the limit cycle of the vdP oscillator calculated with the PTM with $\delta=10^{-100}$ and using Pad\'e approximants of order $[100,100]$. We are working with $400$ digits of accuracy. }
	\bigskip
	\label{table_T_vdP2}
	\begin{center}
		\begin{tabular}{|c|c|}
			\hline
			$\mu$ &   $T$    \\
			\hline
			110	&   178.9303956907033679871688310695657159323417168135937922472711936862634798797065910073778443003100491  \\
			120	& 	195.0285802556076526910160058643286113856413040774148397015025046291085887859870658756179389085971797  \\
			130	& 	211.1307477804160260130832331863590232741382508015255437999265521274563460940912363700317134456024906  \\
			140	& 	227.2362316910533012521444863792345519317926635296215340947212434470538699636331704863156601505403007  \\
			150	& 	243.3445145087953931135502082098788891571273471387131884042823631594633493846569879423539468834013124  \\
			160	& 	259.4551871552089954913517277272206473199985957741243565799737323251409484803982037303780223559037179  \\
			170	& 	275.5679211327523211576511486355320960580031645969284306833900520902209914645207652127999437789348208  \\
			180	& 	291.6824490104429176636024212529204140524163494516962540912121426736499348947208292897874488332050626  \\
			190	& 	307.7985504241069457922015591780920378614606316407770889761198474674473028851653136541693381088651136  \\
			200	& 	323.9160418323323196052744334743514572330591352304211850032764200605474695617827325555219016088358064  \\
			210	& 	340.0347688880652243490443513460132673695170014737344893976990248738694057571245968207911050023418718  \\
			220	& 	356.154600668410400306729858632752533198225570464710167911682149586955622615539789822531926450644064  \\
			230	& 	372.2754252482098363564602268458485149789880169166789785925357810440972561748225119598807497831772816  \\
			240	& 	388.3971462610618011013038198916975395522677722189419414224652127005211074077504912273720584228538532  \\
			250	& 	404.5196801965184163871870483252417534129149893308985855734118257674185692704093695250596637461397278  \\
			260	& 	420.642954253411715287201586295020032714895077340505247244349853711424204153908462757750556288205757  \\
			270	& 	436.7669046183784061997777264506815647938055223685815973519157893021809412437496513849994688098179347  \\
			280	& 	452.8914750730853333239771225564670217954080152594118401826623836805454265359833714379551739495594771  \\
			290	& 	469.0166158581516320832875103415375882683397513540431583325719881015948320585564394337012616182121935  \\
			300	& 	485.1422827394258644411994738128162929528511728967565610619365505973036022082905924814438907274623181  \\
			\hline
			\hline
		\end{tabular}
	\end{center}
	\bigskip\bigskip
\end{sidewaystable}

\begin{sidewaystable}
	\caption{Period of the limit cycle of the vdP oscillator calculated with the PTM with $\delta=10^{-100}$ and using Pad\'e approximants of order $[100,100]$. We are working with $400$ digits of accuracy. }
	\bigskip
	\label{table_T_vdP3}
	\begin{center}
		\begin{tabular}{|c|c|}
			\hline
			$\mu$ &   $T$    \\
			\hline
			310 & 501.2684362351727278043262197886224674534152153985523980152199838682809188502353717678108534815970038 \\
			320 &  517.3950409722499517962169558534433819784295128655886321961149836442728358619817664738133788941608835 \\
			330 &  533.5220651464685439093770993809975226199501092612890489460184777692815090361258658346616492522214782 \\
			340 &  549.6494800676934839949154867656319140696297368261640221561901546724731094473412552942021120600295546 \\
			350 &  565.777259774325139575458418804489197232441836205966285287831574516585916404473258793966102439117355 \\
			360 &  581.9053807049371611599899409528464197025076079369324177055544818049076752372353891929765193185005965 \\
			370 &  598.0338214172741708554742746463809787178465713710050924741281459722903096565639805226055835636146146 \\
			380 &  614.1625623467065316667294887459766934907192942444561368480601789297721379562415973388869487329968018 \\
			390 &  630.2915855977279002588325318270425179584336562242802573981195152721692507775053825636924507305593747 \\
			400 &  646.4208747632589801817235754087546670050667520695370442192859197917172073891282189403517007231212122 \\
			410 &  662.5504147674587935472268172823212596053856889929061113486650534518559143908813267950837851161107143 \\
			420 &  678.6801917284962811716715427761830037224858485711644890650826028465201225367955937090758859653062049 \\
			430 &  694.8101928383406651011117404415935456713454473520735101239397598381973576287877899341686382922655295 \\
			440 &  710.9404062571197622304402500844361021088511568057723287903810936635278920433865554930231362263023076 \\
			450 &  727.0708210199951894034096387862684901871433733460526811373298258977410795037769608257479752148460392 \\
			460 &  743.2014269548306432660248072532951912484203422214028006802759653947115260819969130621543319895466478 \\ 
			470 &  759.3322146091985929587520824066647575473765056190071596350044144396300656133940892613873226111758234 \\
			480 &  775.4631751854931013470343234244302326331407339369619665047832135462539340067800370265770753832435831 \\
			490 &  791.5943004831010125943528547313571368118302188840571420579003377635408654933040581277861246171618416 \\
			500 &  807.7255828467374855142099344594401073186279176290097708714275059764518212981435675295237786722997872 \\
			\hline
			\hline
		\end{tabular}
	\end{center}
	\bigskip\bigskip
\end{sidewaystable}

\begin{sidewaystable}
	\caption{Amplitude of the limit cycle of the vdP oscillator calculated with the PTM with $\delta=10^{-100}$ and using Pad\'e approximants of order $[100,100]$. We are working with $400$ digits of accuracy. }
	\bigskip
	\label{table_A_vdP}
	\begin{center}
		\begin{tabular}{|c|c|}
			\hline
			$\mu$ &   $A$    \\
			\hline
			1   &  2.008619860874843136509640188362640366192026137720714461127703247615558726584991155953057790709097765 \\
			2   &  2.019891384667136024114287489105850930508254812000941664549543712485830284626252034906272251568995642 \\
			3   &  2.023304141682214306513786130455064525739013679872760990999936976560586572275276092802646061447915499 \\
			4   &  2.022962500968811846082061224390385183025813549039480501113893539375551541011962157657937092917287882 \\
			5   &  2.021508061562321323838742024599621246152537461624864225861969635127894986168530050183097939110542784 \\
			6   &  2.019830213891253240477287224737578444042958201799808980661786084367009274112937817502956660484568965 \\
			7   &  2.018215300068836265908942731134283705382531204763517722012146205824350791436123416051201516053594633 \\
			8   &  2.016747479245613790416849586519869058970854114589853082447278090788224517016011905821777373488753969 \\
			9   &  2.015440797962394961832076120470580842408048777323535235111024076919292007549052285584544263913898855 \\
			10  &  2.014285360926405285327639819151920183122053633536581375785514775840505595191629394412431198122197533 \\
			20  &  2.007789970865430297274867350090457336303719272622512696580222467839163820644590094385995046731852068 \\
			30  &  2.005164254840654557051777186680049721988490763259078944821305865733440807262686417365186060483853802 \\
			40  &  2.003789381619443789571203145287844923089135343779111957674707624353600707572222969176109242784488704 \\
			50  &  2.002955933757648402826109718945871764979914720146069491639268919228280941255065893414456892214808004 \\ 
			60  &  2.002401919063881473920962090208714663235156271998258030687422113141268600481080681056348881636454035 \\
			70  &  2.002009596082639609289697462083514119112813772856077413775296699647238878049627998777911871776993943 \\
			80  &  2.001718652710708668905375467432853701724717282116218318894524355022607543603472606814812673333891301 \\
			90  &  2.001495168861862908252915078070973635620733425887828025607985838234406501535795329520921218813904855 \\
			100 &  2.001318681177224161237412765674824826667922262552186082947860663645944718541257075874824291065133597 \\
			\hline
			\hline
		\end{tabular}
	\end{center}
	\bigskip\bigskip
\end{sidewaystable}

\begin{sidewaystable}
	\caption{Amplitude of the limit cycle of the vdP oscillator calculated with the PTM with $\delta=10^{-100}$ and using Pad\'e approximants of order $[100,100]$. We are working with $400$ digits of accuracy. }
	\bigskip
	\label{table_A_vdP2}
	\begin{center}
		\begin{tabular}{|c|c|}
			\hline
			$\mu$ &   $A$    \\
			\hline
			110	&   2.001176150664201236511554898162392329309803631919700088396165021635090622731953166376363185345190468 \\ 
			120 & 	2.001058896554125927029392927438080919331098636531505741935700786658637918519040964812470011465259065 \\ 
			130 & 	2.000960927844888534197421329197213513163533964064997696319356875362523084014342159495491081566374533 \\ 
			140 & 	2.000877983652867070899563966628262022269236536408068634986331923205730266248081399764040778573457873 \\ 
			150 & 	2.000806955271625817278643513939997236295393278928069489496739293779680272206361514498205112393886658 \\ 
			160 & 	2.000745524338127997986533399402526338459501029215022413403802471156496719380286552841303498699166454 \\ 
			170 & 	2.000691928650507049010444693810276888632234923699748845764204325369807162600992668870049753693914348 \\ 
			180 & 	2.000644806183587289954560918219707872421639629764757465033098886830120895062335564168746005183167098 \\ 
			190 & 	2.000603088546078420297503695261372256400644888894879382845043708598328299403205400741349039400898667 \\ 
			200 & 	2.000565926576760211522417824772932917267562790191698912402919850289563593680876332599753138742066905 \\ 
			210 & 	2.000532637350753826167807899544044099402616675472884782815702328994100650744957001382965323763051867 \\ 
			220 & 	2.000502665763518248325354257921241551809517650968823097692036278979440318254163946014190489411599806 \\ 
			230 & 	2.000475556236848568636954324017185605467137320938779282327081023791577576731757080734492140855109316 \\ 
			240 & 	2.000450931578303478694537510744220248976114354365642128210382176047146010114604394305189436058795741 \\ 
			250 & 	2.00042847697772345185029040695739632649701562629941177029801599465199850162343322646453194120400036 \\ 
			260 & 	2.000407927747058093988748823717862181612795541998090284257389144325547078829092172149764902564719009 \\ 
			270 & 	2.000389059824524017306726894783981061575986906859073578895751544091178277158764196266029696601424629 \\ 
			280 & 	2.000371682345317609157227907905005606795933433938383565745122081496027981775545646109723946722435718 \\ 
			290 & 	2.000355631774790119931882114768567816619242450522793582699291719911732054675995403444109331703946892 \\ 
			300	&   2.000340767235358440381118336572999553385449901440431020937886115010068300706803207636643738494352265 \\
			\hline
			\hline
		\end{tabular}
	\end{center}
	\bigskip\bigskip
\end{sidewaystable}

\begin{sidewaystable}
	\caption{Amplitude of the limit cycle of the vdP oscillator calculated with the PTM with $\delta=10^{-100}$ and using Pad\'e approximants of order $[100,100]$. We are working with $400$ digits of accuracy. }
	\bigskip
	\label{table_A_vdP3}
	\begin{center}
		\begin{tabular}{|c|c|}
			\hline
			$\mu$ &   $A$    \\
			\hline
			310 &  2.000326966754318491752882158288726666347822269174555888819685335381219042241539507445589914126673444 \\
			320 &  2.000314124228514365047590787480224800388011045630717020905323049774289130633510670020231543638910683 \\
			330 &  2.000302146951734816161384032376729083130293740082412165493512016110882487677683585416959621768557526 \\
			340 &  2.000290953587329248234887345448502018306154097277200952370316129354246338636069275694592131936921866 \\
			350 &  2.000280472495673520147801804304281450222834146751154194402383792845458390614590018139282388728999713 \\
			360 &  2.000270640346418482727432796549274334807090723809457625301021912279926785202157716023007066231316422 \\
			370 &  2.000261400960778407272014150919879490328851961702490111973205517604442731436470785107417568540835796 \\
			380 &  2.000252704340780169883024954575103776332039887880106187365966367674615954233486090162340690156807203 \\
			390 &  2.000244505851341729254186335289587647349025191125715300019568047920146842751944159064234618604729996 \\
			400 &  2.000236765527963692701424636298007565995335294141201369885707909024705207984905021905264276738379989 \\
			410 &  2.000229447488199892183365006648876507936133576964447728989832214520071784090497965595824113229859935 \\
			420 &   2.000222519429289704999922115937529332436405089602733553210761715084325381901726400869191966041257453 \\
			430  &  2.000215952197659532174091044980369619020676193843196225450138050837614077484865400111023077591425039 \\
			440  &  2.000209719418637855073353468847162792740151388072266145846231279294924400850726940839455543996263235 \\
			450  &  2.000203797176831785484489994634091962887707480896472133778935604952800033427160310818087648760106876 \\
			460  &  2.000198163739300046960156775279678371962466920303239147813726832400200911102449590548664241320247353 \\
			470  &  2.000192799315017326821486381385659974607662167178038199023400878993930245266767520737835363462367545 \\
			480  &  2.000187685845226716315931997510390663860906566981771710897933276683167515206752217716063468330289276 \\
			490  &  2.000182806820173743333076404516509646425847607747067792646395379510237248236860107757251919041250839 \\
			500  &  2.000178147118448723671856525505949565588507795231205946274908215891589874712255970865077462000262459 \\
			\hline
			\hline
		\end{tabular}
	\end{center}
	\bigskip\bigskip
\end{sidewaystable}

\begin{sidewaystable}
\caption{Maximum value of $\dot{x}$ for the limit cycle of the vdP oscillator calculated with the PTM with $\delta=10^{-100}$ and using Pad\'e approximants of order $[100,100]$. We are working with $400$ digits of accuracy. }
\bigskip
\label{table_vmax_vdP}
\begin{center}
\begin{tabular}{|c|c|}
\hline
$\mu$ &   $\dot{x}$    \\
\hline
1   &   2.6784414796069779492758218717041165621077368697191220964124546113071122686772143437815944
\\
2   &   3.8172216408312205563965799781101066245898607996904959302878480206566251386554161251558923
\\
3   &   5.0648773819723163313094116853524203220619739803178721036819802328561052810978640741388417
    \\
4   &   6.3441577400432038683444489495718089498685902242889595120372918990750631357798315295800320
\\
5   &   7.6371588273860254484953749805726801885948506076972919488005689076979283843546367829063331
\\
6   &   8.9379265819413622486569095061412886656746615148297593254782758864921186294387690286246242
\\
7   &   10.243794656977241172415068397372709368729830364880316887707229147049879457120948591090658
\\
8   &   11.553323755040956833345776180236344622718497189835206692188077822200414279169925365982556
\\
9   &   12.865635574375770209066226538411846913192667209823476911233067532624282122142459813282810
\\
10  &   14.180147287056470109063968143632562534202857436059307253774820857605903106639033810558353
\\
20  &   27.390332563171141322681384744174433642333982124083832955063378840095591726423886649706583
\\
30  &   40.653836135205017065709893210160124329955806379520772725310870032465171605910130941182905
\\
40  &   53.939480146757428322281180378086305396823975975294897351542494679584494962135154624339179
\\
50  &   67.237148179968240764712810837764727882207236265536465693912439204011815698169042985428902
\\
60  &   80.542299931770918117476860190131047136151733857026820765008360560464480772050828880468764
\\
70  &   93.852522046193537276316206926104416268177579448390775027959973501001465942801592916712924
\\
80  &   107.16638617332237771851982538617271057517574967536246444479152365251483641747573929637063
\\
90  &   120.48298094138217517882608897256362833109552229133719092225459533635819634172119234500904
\\
100 &   133.80169145553263028831023868627805726314887728137538640478489874691868312722921925081722
\\
\hline
\end{tabular}
\end{center}
\bigskip\bigskip
\end{sidewaystable}

\begin{sidewaystable}
\caption{Maximum value of $\dot{x}$ for the limit cycle of the vdP oscillator calculated with the PTM with $\delta=10^{-100}$ and using Pad\'e approximants of order $[100,100]$. We are working with $400$ digits of accuracy. }
\bigskip
\label{table_vmax_vdP2}
\begin{center}
\begin{tabular}{|c|c|}
\hline
$\mu$ &   $\dot{x}$    \\
\hline
110 &   147.12208452923183079223354494148704870316218792442327102974136328427411539330912071137610
\\
120 &   160.44384427331336092670444558497163203146280727621822158311421948760631510643918746755420
\\
130 &   173.76673374163240752927680778384291805738198058832372192179225854331997267287055093523806
\\
140 &   187.09057097213406649211204994641882735294006592457523986103634256782484946191678590775974
\\
150 &   200.41521341594866362109745804500324331356527133156493638336252591661747429011656568823433
\\
160 &   213.74054747506859932847051940969531349622987851567298175192750066813601687006062717611692
\\
170 &   227.06648126973966070467142510074722986470281460478410567928262460893681984892022145780853
\\
180 &   240.39293951412513904239597521758691504009488921717616045602561106988731904995178999393722
\\
190 &   253.71985980691115194157229177340267325291091272796703094681620418153480326170944646450976
\\
200 &   267.04718989488479834002673374274464706993337572325474692208178275058654460941091737537682
\\
210 &   280.37488562008138856270709478203920101376233653607792481070462817142667813970759023393572
\\
220 &   293.70290935643811499880844866760625529541675548552525479683956772793422567135167155086473
\\
230 &   307.03122880303500190335586310402507162032216417479822183680177835036681285362235565895880
\\
240 &   320.35981604113367281983446853728008884514945216723055112727362380760473281611539018771505
\\
250 &   333.68864678911596726618558582797079510050009698679388087429628011416805583845363693950139
\\
260 &   347.01769980778696146493011278628799152796435071913272132550229342559919895026410513503665
\\
270 &   360.34695642126174726909554725439822485694195246758205536208967730049293082820812183408460
\\
280 &   373.67640012765429281730704326421937148612986460682029446327593779479439738928376880508176
\\
290 &   387.00601628022729728709677301715860802499206616289671859832051152934792256860737660857592
\\
300 &   400.33579182433259013427316975846527931562438027791521849707629780978885026488846159552034
\\
\hline
\hline
\end{tabular}
\end{center}
\bigskip\bigskip
\end{sidewaystable}

\begin{sidewaystable}
\caption{Maximum value of $\dot{x}$ for the limit cycle of the vdP oscillator calculated with the PTM with $\delta=10^{-100}$ and using Pad\'e approximants of order $[100,100]$. We are working with $400$ digits of accuracy. }
\bigskip
\label{table_vmax_vdP3}
\begin{center}
\begin{tabular}{|c|c|}
\hline
$\mu$ &   $\dot{x}$    \\
\hline
310 &   413.66571507890014386006250291257692578257075635215110402035920641848601379112149247928003
\\
320 &   426.99577555377900095286233316020342562536319171461843296415468741232115203698955685014774
\\
330 &   440.32596379614279083992572283454818966379847649093282347590544564733712983239326930531950
\\
340 &   453.65627126061890845953267903252101913196885669749908849995144879267654170297506646452663
\\
350 &   466.98669019890611973039684388515400333114386227041618316698028253113432559419873822267697
\\
360 &   480.31721356549783356906991267038344982568663609377113694924670048035834303819499183430127
\\
370 &   493.64783493679079769122667321628778280687310487121984986466244925956821515082933928703536
\\
380 &   506.97854844137774513451984164556948773123276671664129503113879661135599057205862549645229
\\
390 &   520.30934869973161216566298861834324452117499936926731174319055948823137910566162425916609
\\
400 &   533.64023077181371247846633994461798496480179062908470143012628279727636257504522955478361
\\
410 &   546.97119011139770121185871186421229079972372663431742564709797335625632887296689028790331
\\
420 &   560.30222252610967178010398544821226485793463408448051829657214534480409060442232165315607
\\
430 &   573.63332414235325042824748424156107395867830271862862006430174111368297803846118429787629
\\
440 &   586.96449137442548972469096038693945400277581323438252319611735946295828417937049857184323
\\
450 &   600.29572089724120214994550648282891200261397796961381936807944988451179447473201759040243
\\
460 &   613.62700962217516264917044730584123641961570410927994332350856985868990894808390417740241
\\
470 &   626.95835467560728915698130350144748019910018319274329817901155143002606915486727249704658
\\
480 &   640.28975337981858544006760734119909966859865078487705795822283027038568626497091488716420
\\
490 &   653.62120323593775518061552644205195029590277083980123239603349702479389694488322842403780
\\
500 &   666.95270190868192124866527156295040370726365974421093980155055591304868276944533414166998
\\
\hline
\hline
\end{tabular}
\end{center}
\bigskip\bigskip
\end{sidewaystable}

\begin{sidewaystable}
\caption{$T_{\rm fast}$ for the limit cycle of the vdP oscillator calculated with the PTM with $\delta=10^{-100}$ and using Pad\'e approximants of order $[100,100]$. We are working with $400$ digits of accuracy. }
\bigskip
\label{table_tfast_vdP1}
\begin{center}
\begin{tabular}{|c|c|}
\hline
$\mu$ &   $T_{\rm fast}$    \\
\hline
1   &   2.316802059232911233439124995615143484133199971906039815597790582443718317051451389443049992223249902   \\
2   &   1.777713493687930228693080909813198522206478850618103270642063691764857236876743558361848573069014008   \\
3   &   1.434713670252197367942884953956690225383078933406793439861715958876264146279784216058747272189405444   \\
4   &   1.205488824913180452128110106137254471424619708157288530518049487630065704803466312508798102605123689   \\
5   &   1.042385826116529616294742230604444158424209600220010600236918843458039150955926470937009981808343964   \\
6   &   0.9203711272321695914418075274270152808920816275412675448287713379208258479987592111795952224407453848  \\
7   &   0.8255257047804702162940733671545269671130752310491652286640745426815504146367190814510472226819979058  \\
8   &   0.74957092426707007504677262410273154429702263910323297054904630581557125166834951472313058227776429    \\
9   &   0.6872890373963839619984049707112223821375817647520512801955074242591256485241401206307631421350847956  \\
10  &   0.6352304981733229705221265355959715616294081137472675272451252611531209598529070714713593970734422273  \\
20  &   0.3700517114131607725303032457137734456674085058123833103466836467601039358776259199881695557925490004  \\
30  &   0.2661647559294058983195734514406499895060505427128870340111393756172988422864389378985703508894970482  \\
40  &   0.2097544033992348395743063069335438114523348211501552182331645069578351542964384111384370423210270942  \\
50  &   0.1740116387481188338090420640851550289909218423568297625120289676633181910726332961613471976694907271  \\
60  &   0.1492027651949420262833172732014503662767979817121420229861714087666780353540376812030497413358159289  \\
70  &   0.1309098442211953756863757835898331692550660308855217594274411698547075429102098309476668550945587847  \\
80  &   0.116827095590314924552874784608225378747676441100706416021203117426544677542541060639158502067443325   \\
90  &   0.1056291253057036502848574109204996528410558751987790010487601928892872187049639691283217588080588711  \\
100 &   0.0964979826308564096212857964676339416835886778477010364996693042574994139792286310313956590068728237  \\
\hline
\hline
\end{tabular}
\end{center}
\bigskip\bigskip
\end{sidewaystable}

\begin{sidewaystable}
\caption{$T_{\rm fast}$  for the limit cycle of the vdP oscillator calculated with the PTM with $\delta=10^{-100}$ and using Pad\'e approximants of order $[100,100]$. We are working with $400$ digits of accuracy. }
\bigskip
\label{table_tfast_vdP2}
\begin{center}
\begin{tabular}{|c|c|}
\hline
$\mu$ &   $T_{\rm fast}$    \\
\hline
110 &   0.08890054758676575739933973795461472571207215224321633559263795073064215877018871310561743652508994131 \\
120 &   0.08247395265540504975817853948585174758150519928167544215849240119025223492892223674842377763542014062 \\
130 &   0.07696234390911981091787943357987737911010769492576336146688799798092560998947686990367596754362940068 \\
140 &   0.07217995977427006115215892646685531792135953477586364372803009257513850194981792840956772362097937678 \\
150 &   0.06798855386204053436484005954152793414022905944616926081114782953557145848352220682620795863860553181 \\
160 &   0.06428304269112547493553633342667431331510680493475639546043965692234626152760227195632052010919849036 \\
170 &   0.06098207541842125986517863159899378549787830274391685384632849817699680525150109507492409863966865421 \\
180 &   0.05802165854690050579029786322928208924263531735443867336563163187094654481754243056683610198108244012 \\
190 &   0.05535073760764241159312602821530074826260753925240320135601788109002056506492112473591755632207775979 \\
200 &   0.05292806751914627672369764488005262035092004955655317461386711862058998586114511207685822209148608999 \\
210 &   0.05071995244252124024822468750289744519548484018179146143096773238567171286751074069361155486990566307 \\
220 &   0.04869858512094769787934572484352777196978234390791729822803649693412768614896995792343554985673148832 \\
230 &   0.04684080761087943783108571666495809425481768773421005944706212170243143812056243673571390912623301244 \\
240 &   0.04512717342011540421308814427488347094528005739164059418133552990569805541148529373775867221828625996 \\
250 &   0.0435412286543546440596761583155298283755490460653916431548799547151224985962877938535920475406579617  \\
260 &   0.04206895459576740421710531067405207141375135100376591928898349852817095820564931607592108099677060587 \\
270 &   0.04069833084088671708323401020782956111601149570942363800443461541932786695083704494359750734273017557 \\
280 &   0.03941898956049937214356947610386262435988277990111656015873480174963487348240238455345853208312462283 \\
290 &   0.03822193939693922485234190703958107015824523523670799072860870245129122955286855874421375093308360217 \\
300 &   0.03709934312545045421570367284167161358297777705296003272965967121356923310470068787691652535451559648 \\
\hline
\hline
\end{tabular}
\end{center}
\bigskip\bigskip
\end{sidewaystable}

\begin{sidewaystable}
\caption{$T_{\rm fast}$  for the limit cycle of the vdP oscillator calculated with the PTM with $\delta=10^{-100}$ and using Pad\'e approximants of order $[100,100]$. We are working with $400$ digits of accuracy. }
\bigskip
\label{table_tfast_vdP3}
\begin{center}
\begin{tabular}{|c|c|}
\hline
$\mu$ &   $T_{\rm fast}$    \\
\hline
310 &   0.03604433721864130813034264003643260999412566196618118211950008926088598224538898025791531658193299878 \\
320 &   0.03505088435774031663857187763770377083965546689994104783356081926215403434470014976032877274265438587 \\
330 &   0.03411365206142391990236395378981995808342130711134988235286132730153509728758940277985447487918791224 \\
340 &   0.03322791217730221673267326332685888163020191299551568748912807007781543147207252145833064350196303277 \\
350 &   0.03238945715802491863891844940561590977740085370411422115400346724809306483970231183399627269635960467 \\
360 &   0.03159452993197212438812964993613476629230199085396150100024210491270676405298246323326916034406963885 \\
370 &   0.03083976485440747000965403995981194099857709035389994235386059161641087972371636570117130813121993403 \\
380 &   0.03012213774368818252377085474158552574177084607189769933761124184139336478884700098564121681769864462 \\
390 &   0.02943892340829243428612033647353043304552570622568968755781411947036265728285154925599885264302368431 \\
400 &   0.02878765938294597163977772635237857912179269891946167327360827043623908429488566733392315057343383515 \\
410 &   0.02816611483727510730736941616758955385192552639383640996814464895663041650474225258042747239113474034 \\
420 &   0.0275722638139629257921487838486741719455881755522336748354394044824890886624781423742017181088626046  \\
430 &   0.02700426210714431260573257377310805030582255954475488634956230346854255949376924032736101529681147961 \\
440 &   0.02646042721464038513881351777829810909181927516349626287694729867135912727879189903480935294231397692 \\
450 &   0.02593922089635869348793883695531127519692230290938982845584726724201488573803576871152073423283386174 \\
460 &   0.0254392339509354677837973397939151219554958513522711867201408831668325095338373015550424779983068977  \\
470 &   0.02495917288744291768778307266904557735895293522819018974877863744278277838685754313803582857624017321 \\
480 &   0.02449784822180383531876698346992944286160841740021207249533097856794421880505320070948742930404486578 \\
490 &   0.02405416417084288660585935815023390854958454164716020112494014453659348770394492765901134644828355651 \\
500 &   0.02362710955253510555549371153382972768459511168489054990560767836010998889848894002345766774852635714 \\
\hline
\hline
\end{tabular}
\end{center}
\bigskip\bigskip
\end{sidewaystable}

\begin{sidewaystable}
\caption{$-\lambda_2/\mu$ for the limit cycle of the vdP oscillator calculated with the PTM with $\delta=10^{-100}$ and using Pad\'e approximants of order $[100,100]$. We are working with $400$ digits of accuracy. }
\bigskip
\label{table_Lyap_vdP1}
\begin{center}
\begin{tabular}{|c|c|}
\hline
$\mu$ &   $-\lambda_2/\mu$    \\
\hline
1   &   1.059376994841854892700403949338049481957020648372259955756465472536805561282965199891004946447931001   \\
2   &   1.191280245015429832634801219084992602555987495901929785416197241010264364520523313504648975995014355   \\
3   &   1.313036827642448510307624855182919001385607909683629066933147586842521678962347520915813767460325009   \\
4   &   1.404448873378790759718579148766875456066757513825968814919695490152201505630332608599767178554790909   \\
5   &   1.471758889227520934456604638534863976948547891972705047241953908170682903815834983042903928663031673   \\
6   &   1.52214411621512969910941982913582749376733959428564402783933515736610345365499037059894456466232475    \\
7   &   1.560716682574959343937952193600900451400925042380450856598239957582346950627536220378464844395001546   \\
8   &   1.590898442984244587906591546650111128409345173903300768848151198927108921400039888846034941782889802   \\
9   &   1.614984707150109742044802782906511875013313466132843411125073706400635468630590270929565212004124396   \\
10  &   1.634543340782244633737452978422855845593454956242903739095118631368197795051991852882706070372940806   \\
20  &   1.722439943215048604560426095471226646618067577034101988905091127372192868333411180079223220912250248   \\
30  &   1.749480232379767914492545039837800508276854829492984632612447597022648010708704161729233330759058748   \\
40  &   1.761859321198200014299605833239624080908380527784372355531263737247914081724045460050488339278384774   \\
50  &   1.7687483241232140509598593651478800697087563956447734439340080354348283569383416391856046766055076 \\
60  &   1.773056829698572169218660456801393137857290498848373576228815536080871261177198135796328822000174224   \\
70  &   1.775969601945597352354204099935671239834793159462966113706958208916076933541459143558966677072888234   \\
80  &   1.778051519226189073707012901163427897456102909840577364334639275960877980140044867184742915476446536   \\
90  &   1.779603106434321333179373250845245933723786540633952069391023609779631970199986673322440731457916575   \\
100 &   1.780797717789814324505111213833634637169341499665322580329707533781620319604374725252941547690871148   \\
\hline
\hline
\end{tabular}
\end{center}
\bigskip\bigskip
\end{sidewaystable}

\begin{sidewaystable}
\caption{$-\lambda_2/\mu$ for the limit cycle of the vdP oscillator calculated with the PTM with $\delta=10^{-100}$ and using Pad\'e approximants of order $[100,100]$. We are working with $400$ digits of accuracy. }
\bigskip
\label{table_Lyap_vdP2}
\begin{center}
\begin{tabular}{|c|c|}
\hline
$\mu$ &   $-\lambda_2/\mu$    \\
\hline
110 &   1.78174177611711785619288827587863025730938224399643781920953899565850647339299267453047447341775541    \\
120 &   1.782503929370829416107244800115417078891648329444677706117929911990126054519944306605824388684606679   \\
130 &   1.783130280000057627766732659336996063047019701679233093379142494238534719463590010406658115773094781   \\
140 &   1.783652843384642536212044166926501834296201376239104998411362463090874637778940709756535630353929235   \\
150 &   1.784094489474281384775781089735677536861981176623433450021721638599529601577652025870138700433793632   \\
160 &   1.784471955031130100135393545267883564339384278665052177577483323688834746468818585590506758351012645   \\
170 &   1.784797747389768931746889300509792863352538938138177427762437380233887329032634672652905655073469093   \\
180 &   1.78508138575536651357316878455041473733113048989789314914889887328813270301242215789153109170061227    \\
190 &   1.785330232856269839045173716801627311313602067838367019361120762648997832671422501488887959497406314   \\
200 &   1.785550065617615830325000436866601061088945887796444415850663749442755893628189952707295809058828599   \\
210 &   1.785745475139290187619060100732447627533747529916540590540065273981802480093244046733413101369089923   \\
220 &   1.785920152389285458758187025156333614517370214185824788315531316888193000089223659347521009976936451   \\
230 &   1.786077095764333480508413209255625010100723267896790757021380628922002156137142644043780069388592768   \\
240 &   1.786218764218880432666141602650328455887770282492350624224686466843520235974538789701816059957864711   \\
250 &   1.786347191822507657860263018010957365719695877575972310258294366824673573470078599821537278119116799   \\
260 &   1.786464074557974685652103827166213869044666083595504607101598804109671770927137939768306998842820506   \\
270 &   1.786570836856577956085807542528207289484483680724589544858875935189026812384711094819648946604728776   \\
280 &   1.786668683149796316538855065852865869474244951373266320571991388048047267435701299541665384552485657   \\
290 &   1.78675863820775447191896368443481200475150287205048794648255403233723959062692099705715996617525039    \\
300 &   1.786841578993139423476252515213511891583858172664916064973795221667260472689934421913201377666690037   \\
\hline
\hline
\end{tabular}
\end{center}
\bigskip\bigskip
\end{sidewaystable}

\begin{sidewaystable}
\caption{$-\lambda_2/\mu$ for the limit cycle of the vdP oscillator calculated with the PTM with $\delta=10^{-100}$ and using Pad\'e approximants of order $[100,100]$. We are working with $400$ digits of accuracy. }
\bigskip
\label{table_Lyap_vdP3}
\begin{center}
\begin{tabular}{|c|c|}
\hline
$\mu$ &   $-\lambda_2/\mu$    \\
\hline
310 &   1.78691826002927673597152003923831414730216898318076648842973224979232446064090615790200414200407189    \\
320 &   1.786989333762945593166104931964156491048179082572303192393045870355323466057570745057405434255062053   \\
330 &   1.787055367030217304377336997180721396384405647716801198833480512776829400220010871093591890001124151   \\
340 &   1.787116854463042554977194105150964006054424937055872986748162774681485479221412973068857837595516424   \\
350 &   1.787174229475594962227081379512504657482863475893642504026391167196886874692111504303315697767991469   \\
360 &   1.787227873321973688424660568203179763471850852741150371872427173568253057780052237610503129306600055   \\
370 &   1.787278122606506049350383004355189359311694781240185295918951241904814428491761244116367604332245166   \\
380 &   1.787325275544539468140766095549073505460457138093721401903077170426774143417520637753305576399661407   \\
390 &   1.787369597208141295020175549763917328519462963091142673040983315391092792637724438750655705967000721   \\
400 &   1.787411323942417466270049947468143290908933659128835974986432456181534171228523332673112879429068705   \\
410 &   1.787450667100508488455778812942116424086362427920748538023274495073201991195781681181840460396373946   \\
420 &   1.78748781621601263121176561071223007699482711134378573919500300848142429586114030265442062286400958    \\
430 &   1.787522941708622460410273191372541462738456114313258451946870938679693980639889010159500063230616684   \\
440 &   1.787556197200656147093695332892021139742119945826491904716798402091028061627663758359330359483157125   \\
450 &   1.787587721507806599779528997754031252663504134592667153611703894251799888412617813802116762523874582   \\
460 &   1.787617640355979995510073549485759603908037526063997057738198637092526633367604113210622984058074461   \\
470 &   1.787646067866913179064815378444754022460060691477132436954868200579432038066252269583321241201568027   \\
480 &   1.787673107847859245009140585644646478616157542233691654721331591922088178138574503318819354650598944   \\
490 &   1.78769885491463745032907833725229437884488958530445603843186894595499362688388494704563626528344106    \\
500 &   1.787723395472467251065110696980644664108394193152283582383655991387804113862476695523341720712451181   \\
\hline
\hline
\end{tabular}
\end{center}
\bigskip\bigskip
\end{sidewaystable}


\begin{thebibliography}{}
	
\bibitem{Deprit66} Deprit, André, and R. V. M. Zahar, "Numerical integration of an orbit and its concomitant variations by recurrent power series." Zeitschrift für angewandte Mathematik und Physik ZAMP 17.3 (1966): 425-430.
\bibitem{Corliss82} Corliss, George, and Y. F. Chang, "Solving ordinary differential equations using Taylor series." ACM Transactions on Mathematical Software (TOMS) 8.2 (1982): 114-144.
\bibitem{Chang94} Chang, Y. F., and George Corliss, "ATOMFT: solving ODEs and DAEs using Taylor series." Computers \& Mathematics with Applications 28.10-12 (1994): 209-233.
\bibitem{Jorba05} Jorba, Àngel, and Maorong Zou, "A software package for the numerical integration of ODEs by means of high-order Taylor methods." Experimental Mathematics 14.1 (2005): 99-117.
\bibitem{Barrio05} Barrio, R., F. Blesa, and M. Lara, "VSVO formulation of the Taylor method for the numerical solution of ODEs." Computers \& mathematics with Applications 50.1-2 (2005): 93-111.
\bibitem{Barrio05b} Barrio, Roberto, "Performance of the Taylor series method for ODEs/DAEs." Applied Mathematics and Computation 163.2 (2005): 525-545.
\bibitem{Barrio11}  Abad, Alberto, Roberto Barrio, and Angeles Dena, "Computing periodic orbits with arbitrary precision." Physical review E 84.1 (2011): 016701.
\bibitem{Barrio12} Abad, Alberto, et al, "Algorithm 924: TIDES, a Taylor series integrator for differential equations." ACM Transactions on Mathematical Software (TOMS) 39.1 (2012): 1-28.	

\bibitem{VanderPol28} van Der Pol, Balth, and Jan van Der Mark. "LXXII. The heartbeat considered as a relaxation oscillation, and an electrical model of the heart." The London, Edinburgh, and Dublin Philosophical Magazine and Journal of Science 6.38 (1928): 763-775.
\bibitem{Ginoux12} J.M. Ginoux and C. Letellier, ``van der Pol and the history of relaxation oscillations: Toward the emergence of a concept", Chaos 22, 023120 (2012)
\bibitem{Amore18} Amore, Paolo, John P. Boyd, and Francisco M. Fernández. "High order analysis of the limit cycle of the van der Pol oscillator." Journal of Mathematical Physics 59.1 (2018): 012702
	
	
		
\bibitem{Krogdahl60} Krogdahl, Wasley S. "Numerical solutions of the van der Pol equation." Zeitschrift für angewandte Mathematik und Physik ZAMP 11.1 (1960): 59-63.
\bibitem{Urabe60} Urabe, Minoru. "Periodic Solutions of van der Pol's Equations with Large Damping Coefficient $\lambda= 0 \sim 10$." IRE Transactions on Circuit Theory 7.4 (1960): 382-386.
\bibitem{Ponzo65b} Ponzo, P., and N. Wax. "On the periodic solution of the van der Pol equation." IEEE Transactions on Circuit Theory 12.1 (1965): 135-136.
\bibitem{Zonneveld66} Zonneveld, Jacob Anton. "Periodic solutions of the van der Pol equation." Indag. Math 28 (1966): 620-622.
\bibitem{Clenshaw66} Clenshaw, C. W. "The solution of van der Pol's equation in Chebyshev series." Numerical Solution of Nonlinear Differential Equations (Greenspan, D., Hrsg.), S (1966): 55-63.
\bibitem{Greenspan72} Greenspan, Donald. "Numerical approximation of periodic solutions of van der Pol's equation." Journal of Mathematical Analysis and Applications 39.3 (1972): 574-579.
	

\bibitem{Wanner96} Wanner, Gerhard, and Ernst Hairer. Solving ordinary differential equations II. Vol. 375. Springer Berlin Heidelberg, 1996.
\bibitem{Anne04} Kværnø, Anne. "Singly diagonally implicit Runge–Kutta methods with an explicit first stage." BIT Numerical Mathematics 44.3 (2004): 489-502.
\bibitem{Eriksson04} Eriksson, Kenneth, Claes Johnson, and Anders Logg. "Explicit time-stepping for stiff ODEs." SIAM Journal on Scientific Computing 25.4 (2004): 1142-1157.
\bibitem{Fazio08} Fazio, Riccardo. "Numerical scaling invariance applied to the van der Pol model." Acta Applicandae Mathematicae 104.1 (2008): 107-114.
\bibitem{Boom18} Boom, Pieter D., and David W. Zingg. "Optimization of high-order diagonally-implicit Runge–Kutta methods." Journal of Computational Physics 371 (2018): 168-191.

	
\bibitem{Haag43} Haag, Jules. "\'Etude asymptotique des oscillations de relaxation (suite et fin)." Annales scientifiques de l'\'Ecole Normale Sup\'erieure 60 (1943): 65-111; 
\bibitem{Haag43b} Haag, Jules. "\'Etude asymptotique des oscillations de relaxation." Annales scientifiques de l'\'Ecole Normale Supérieure 60 (1943): 289-289; 
\bibitem{Haag44} Haag, Jules. "Exemples concrets d'\'etude asymptotique d'oscillations de relaxation." Annales scientifiques de l'\'Ecole Normale Supérieure 61 (1944): 73-117 
\bibitem{Dor47} A. A. Dorodnitsyn, “Asymptotic Solution of the van der Pol Equation,” Prikl. Mat. Mekh. 11, 313–328 (1947)
[translated to English in "Asymptotic solution of van der Pol's equation". No. 88. American Mathematical Society, 1953.
\bibitem{Dor52} A.~A.~Dorodnitsyn, Asymptotic laws of distribution of the characteristic values for certain special forms of differential equations of the second order, Uspekhi Mat. Nauk (1952), 7, 6 (52), 3--96
\bibitem{Urabe63} Urabe, Minoru. "Numerical study of periodic solutions of the van der Pol equation." International Symposium on Nonlinear Differential Equations and Nonlinear Mechanics. Academic Press, 1963.
\bibitem{Ponzo65} Ponzo, Peter J., and Nelson Wax. "On certain relaxation oscillations: Asymptotic solutions." Journal of the Society for Industrial and Applied Mathematics 13.3 (1965): 740-766.
\bibitem{OMalley68} O'Malley Jr, Robert E. "Topics in singular perturbations." Advances in Mathematics 2.4 (1968): 365-470.
\bibitem{MacGillivray83a} MacGillivray, A. D. "On the leading term of the inner asymptotic expansion of van Der Pol’s equation." SIAM Journal on Applied Mathematics 43.3 (1983): 594-612.
\bibitem{MacGillivray83b} MacGillivray, A. D. "On the leading term of the outer asymptotic expansion of van Der Pol’s equation." SIAM Journal on Applied Mathematics 43.6 (1983): 1221-1239.
\bibitem{Kerimov11} Kerimov, Movlud Kerimovich. "Special functions in Academician AA Dorodnicyn’s scientific legacy: Special functions associated with the van der Pol equation." Computational Mathematics and Mathematical Physics 51.5 (2011): 781-802.

\bibitem{Grasman05} Grasman, Johan, Ferdinand Verhulst, and Shagi‐Di Shih. "The Lyapunov exponents of the Van der Pol oscillator." Mathematical methods in the applied sciences 28.10 (2005): 1131-1139.


	
\bibitem{Deprit67} A. Deprit and A. Rom, "Asymptotic Representation of the Cycle of van der Pol's Equation for Small Damping Coefficients", ZAMP, 18 (1967), 736-747
\bibitem{Deprit68} A. Deprit and A. Rom, Lindstedt's series on a computer, The astronomical journal, 73 (1968), 210-213
\bibitem{Deprit79} A. Deprit and D. S. Schmidt, Exact coefficients of the limit cycle in the van der Pol's equation, Journal of Research of the National Bureau of standards, 84 (1979), 293-298
\bibitem{Andersen82} Andersen, C. M., and James F. Geer. "Power series expansions for the frequency and period of the limit cycle of the van der Pol equation." SIAM Journal on Applied Mathematics 42.3 (1982): 678-693.
\bibitem{Dadfar84} Dadfar, Mohammad B., James Geer, and Carl M. Andersen. "Perturbation analysis of the limit cycle of the free van der Pol equation." SIAM Journal on Applied Mathematics 44.5 (1984): 881-895.
\bibitem{Buonomo98} Buonomo, Antonio. "The periodic solution of van der Pol's equation." SIAM Journal on Applied Mathematics 59.1 (1998): 156-171.
\bibitem{Amore04} Amore, Paolo, and Héctor Montes Lamas. "High order analysis of nonlinear periodic differential equations." Physics Letters A 327.2-3 (2004): 158-166.
\bibitem{Suetin12} S.P. Suetin, Numerical Analysis of Some Characteristics of the Limit Cycle of the Free van der Pol Equation, Proceedings of the Steklov Institute of Mathematics, 2012, Vol. 278, Suppl. 1, pp. S1–S54.
	
	
	
\bibitem{Odani00} Odani, Kenzi. "On the limit cycle of the Liénard equation." Archivum mathematicum 36.1 (2000): 25-31.
\bibitem{Lijun15} Lijun, Yang, and Zeng Xianwu. "An upper bound for the amplitude of limit cycles in Liénard systems with symmetry." Journal of Differential Equations 258.8 (2015): 2701-2710.
\bibitem{Turner15} Turner, Norman, Peter VE McClintock, and Aneta Stefanovska. "Maximum amplitude of limit cycles in Liénard systems." Physical Review E 91.1 (2015): 012927.
\bibitem{Cao17} Cao, Yuli, and Changjian Liu. "The estimate of the amplitude of limit cycles of symmetric Liénard systems." Journal of Differential Equations 262.3 (2017): 2025-2038.
\bibitem{Ignatev17} Ignat’ev, A. O. "Estimate for the amplitude of the limit cycle of the Liénard equation." Differential equations 53.3 (2017): 302-310.
	
		
\bibitem{Amore16} Amore, Paolo, et al. "High order eigenvalues for the Helmholtz equation in complicated non-tensor domains through Richardson extrapolation of second order finite differences." Journal of Computational Physics 312 (2016): 252-271.
\bibitem{Amore19} Amore, Paolo, John P. Boyd, and Natalia Tene Sandoval. "Isospectral heterogeneous domains: A numerical study." Journal of Computational Physics: X 1 (2019): 100018.
\bibitem{Amore19b} Amore, Paolo, and Martin Jacobo. "Thomson problem in one dimension: Minimal energy configurations of N charges on a curve." Physica A: Statistical Mechanics and its Applications 519 (2019): 256-266.
	
\bibitem{Kuttler84} Kuttler, James R., and Vincent G. Sigillito. "Eigenvalues of the Laplacian in two dimensions." Siam Review 26.2 (1984): 163-193.
\bibitem{Jones17} Jones, Robert Stephen. "Computing ultra-precise eigenvalues of the Laplacian within polygons." Advances in Computational Mathematics 43.6 (2017): 1325-1354.	
\bibitem{Gordon92} Gordon, Carolyn, David L. Webb, and Scott Wolpert. "One cannot hear the shape of a drum." Bulletin of the American Mathematical Society 27.1 (1992): 134-138.

\bibitem{szhorvat} Szabolcs Horvát, "LaTeX typesetting in Mathematica",  http://szhorvat.net/pelican/latex-typesetting-in-mathematica.html
\bibitem{wolfram} Wolfram Research, Inc., Mathematica, Version 12.3.1, Champaign, IL (2021).
	
\end{thebibliography}
\end{document}